\documentclass[aps,prl,reprint,superscriptaddress,nofootinbib,floatfix]{revtex4-2}
\usepackage[T1]{fontenc}
\usepackage{amsmath,amssymb,bm}
\usepackage{graphicx}
\usepackage{xcolor}
\usepackage{booktabs}
\usepackage{hyperref}
\hypersetup{colorlinks=true,citecolor=blue,linkcolor=blue,urlcolor=blue}

\newcommand{\ThThree}{\ensuremath{{}^{229}\mathrm{Th}^{3+}}}
\newcommand{\ket}[1]{\lvert#1\rangle}
\newcommand{\bra}[1]{\langle#1\rvert}

\newcommand{\paperTitle}{Probing the Nuclear E2 Transition Amplitude in
  \texorpdfstring{$^{229}$Th}{229Th} via Quantum Interference}
\newcommand{\letterheading}[1]{\textit{#1}---}
\newcommand{\NuclearMOneMuN}{0.7992}

\newcommand{\ThThreeThetaPlus}{-53.50}
\newcommand{\ThThreeThetaMinus}{-39.24}
\newcommand{\ThThreePeakMs}{0.886}

\newcommand{\ThThreePeakPct}{99.93}

\newcommand{\IllustrativeAngleSigmaDeg}{0.5}
\newcommand{\IllustrativeFieldSigmaPct}{1}
\newcommand{\IllustrativeRMESigmaPct}{3}
\newcommand{\IllustrativePhaseSigmaDeg}{5}
\newcommand{\ThThreePhaseSigmaPlusPct}{1.03}
\newcommand{\ThThreePhaseSigmaMinusPct}{0.25}
\newcommand{\ThThreeAngleSigmaPlusPct}{11.92}
\newcommand{\ThThreeFieldSigmaOnePlusPct}{6.53}
\newcommand{\ThThreeFieldSensitivityPlus}{6.53}
\newcommand{\ThThreeSigmaThreePlusPct}{7.06}

\newcommand{\ThThreeAngleSigmaMinusPct}{4.21}
\newcommand{\ThThreeFieldSigmaOneMinusPct}{2.37}
\newcommand{\ThThreeFieldSensitivityMinus}{2.37}
\newcommand{\ThThreeSigmaThreeMinusPct}{4.54}

\usepackage{placeins}
\begin{document}
\raggedbottom
\title{\paperTitle}
\author{Shi-Cheng Yu}
\affiliation{Center for Theoretical Physics, School of Physics and Optoelectronic Engineering, Hainan University, Haikou 570228, China}
\author{Zhen-Xiang Zhong}
\email{zxzhong@hainanu.edu.cn}
\affiliation{Center for Theoretical Physics, School of Physics and Optoelectronic Engineering, Hainan University, Haikou 570228, China}
\author{Xin Tong}
\email{tongxin@wipm.ac.cn}
\affiliation{Innovation Academy for Precision Measurement Science and Technology, Chinese Academy of Sciences, Wuhan 430071, China}
\author{Cheng-Bin Li}
\email{cbli@apm.ac.cn}
\affiliation{Innovation Academy for Precision Measurement Science and Technology, Chinese Academy of Sciences, Wuhan 430071, China}

\begin{abstract}
The nuclear E2 transition amplitude of $^{229}$Th probes nuclear deformation
and configuration mixing, yet remains poorly constrained by measurements dominated
by M1 radiation. We propose to determine its magnitude and sign relative to the
nuclear M1 transition amplitude through quantum interference in $^{229}$Th$^{3+}$.
Two coherent laser fields couple a superposition of initial hyperfine states to a
common nuclear isomeric state through the electronic shell. Varying the mixing
angle controls interference between the excitation pathways, allowing the signed
E2/M1 transition amplitude ratio to be extracted using the calibrated laser-field
amplitude ratio, without absolute intensity calibration. An independent M1
reference then determines the E2 magnitude. Dynamical calculations support the
feasibility of the scheme, predicting appreciable nuclear excitation with
milliwatt laser power on millisecond timescales. We evaluate the extraction
sensitivity to experimental and atomic-structure uncertainties. The approach
extends to other $^{229}$Th ions, whose rich level structures offer a broader
selection of transitions and promising routes to suppress systematic uncertainties.
\end{abstract}
\maketitle

The exceptionally low energy of the $^{229}$Th
clock transition has enabled direct laser excitation
\cite{Tiedau2024,Elwell2024,ZhangFilms2024,Elwell2025,Morawetz2026}
and precision spectroscopy in $^{229}$Th-doped materials
\cite{Zhang2024,Higgins2025,Ooi2026,ToscaniClock2026,HuangClock2026}.
Its sensitivity to variations of the fine-structure constant also makes it
a probe of fundamental physics
\cite{Flambaum2006,Fadeev2020,Yamaguchi2024,Beeks2025}.
These advances motivate a detailed understanding of the structure of the nuclear ground and isomeric states.

The nuclear electric-quadrupole (E2) transition amplitude provides a probe of their
structure. In collective models of $^{229}$Th, rotations and quadrupole--octupole
vibrations of the nuclear core couple to the unpaired neutron
\cite{Ruchowska2006,Minkov2021}.
The leading collective E2 operator
depends directly on the quadrupole deformation coordinate
\cite{Minkov2017}. Octupole correlations
and particle--core coupling modify the nuclear wave functions and the
mixing of intrinsic configurations \cite{Minkov2021}.
These effects make the E2 transition amplitude a sensitive test of nuclear
models describing deformation and configuration mixing.
Accurate knowledge of the E2 transition amplitude would also improve
quantitative predictions of nuclear hyperfine mixing in $^{229}$Th ions by
providing a key nuclear parameter
\cite{Zhang2024Nonlinear,Zhou2026NHM}.

Existing measurements, however, leave this E2 transition amplitude poorly constrained.
At the approximately $8.36$-eV isomeric transition, the estimated free-space E2
radiative rate is about nine orders of magnitude smaller than the magnetic-dipole
(M1) radiative rate \cite{Bilous2018}. Radiative-lifetime and fluorescence measurements
therefore primarily constrain the M1 transition strength
\cite{Kraemer2023,Zhang2024}. Nuclear models and indirect inferences from
higher-lying rotational bands yield different estimates of the E2 transition strength
\cite{Ruchowska2006,Tkalya2015,TkalyaErratum2017,Minkov2017,Minkov2021,Zhang2022,Chen2025PSM}.
The nearly fivefold spread in downward E2 transition strengths, from $8.74$ to
$42.9$ Weisskopf units (W.u.) \cite{Chen2025PSM,Minkov2021}, motivates an independent
measurement of the nuclear E2 transition amplitude.

Transition strengths alone do not determine the relative sign of the M1 and E2
transition amplitudes. Interference between these amplitudes makes photon angular
distributions and angular correlations sensitive to their relative sign,
providing a basis for nuclear multipole-mixing analyses \cite{Krane1974,Smith2019GRIFFIN}.
For the clock transition in $^{229}$Th, however, the weak E2 radiative
contribution makes determining the relative sign from these angular observables
challenging. Electronic bridge excitation offers a route to probing M1--E2
interference through the electronic shell. In hyperfine-electronic-bridge (HEB) excitation,
the hyperfine interaction couples electronic excitation to the nuclear
transition \cite{Porsev2010Th3,PorsevBridge2010,Wang2024}.
The effective HEB amplitude contains nuclear M1 and E2
contributions weighted by electronic matrix elements and intermediate-state
energy denominators. For selected transitions in \ThThree{}, the effective
E2 coupling can be comparable to or exceed the effective M1 coupling
\cite{Bilous2018}.

In this Letter, we propose to extract the nuclear E2 transition amplitude
and its sign relative to the nuclear M1 transition amplitude by controlling quantum
interference in \ThThree{}. A microwave field prepares a coherent
superposition of two hyperfine states of the metastable electronic
$7s_{1/2}$ level, with the nucleus remaining in its ground state.
Two coherent laser fields couple these states to a common final state of the
nuclear isomer. Varying the mixing angle of the initial superposition
controls the relative weights of two excitation pathways with different
relative M1 and E2 contributions. Fitting the excitation signal as a
function of the mixing angle locates the interference minimum.
Its position, together with calculated electronic coefficients and the
calibrated laser-field amplitude ratio, yields the signed E2/M1 transition amplitude
ratio without absolute laser intensity calibration. Combining this ratio with an
independent M1 reference determines the E2 magnitude and its relative sign. Our calculations predict
appreciable nuclear excitation on millisecond timescales with milliwatt
laser power, and we evaluate extraction uncertainties arising from angle
measurements, laser-field amplitude calibration, electronic matrix elements,
and the relative phase between
the excitation pathways.

\begin{figure}[!tbp]
 \includegraphics[width=\columnwidth]{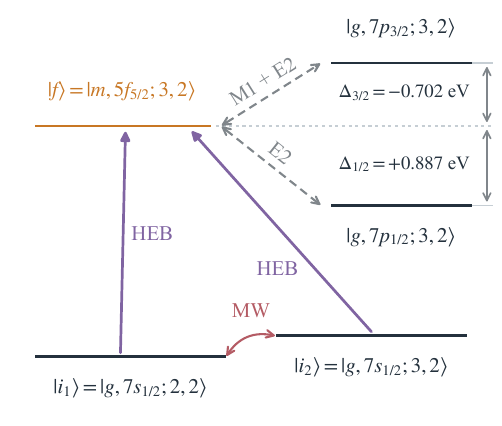}
 \caption{HEB excitation scheme in \ThThree{}. The initial states
$\ket{i_1}=\ket{g,7s_{1/2};2,2}$ and
$\ket{i_2}=\ket{g,7s_{1/2};3,2}$ are prepared in the coherent
superposition of Eq.~\eqref{eq:input} via microwave (MW) coupling.
Purple solid arrows denote effective HEB couplings to the common final
state $\ket f=\ket{m,5f_{5/2};3,2}$. Gray dashed arrows represent
hyperfine coupling between $\ket f$ and the $7p$ states through the
indicated nuclear multipoles. Vertical arrows indicate the signed energy
mismatches $\Delta_n=E_f-E(g,n;3)$. Optical and initial-state hyperfine
intervals are not to scale.}
 \label{fig:platforms}
\end{figure}

\letterheading{HEB excitation}We denote a nuclear--electronic state by
$\ket{\nu,e;F,M}$, where $\nu=g,m$ labels the nuclear ground and
isomeric states, respectively, $e$ labels the electronic state, and $F$ and $M$
are the total angular momentum and magnetic quantum numbers.
For the two initial states shown in Fig.~\ref{fig:platforms}, the prepared
superposition in the rotating frame of the two laser fields is
\begin{equation}
 \ket{\psi_i(\theta)}=\sin\theta\ket{i_1}+\cos\theta\ket{i_2}.
 \label{eq:input}
\end{equation}
A microwave pulse controls the mixing angle $\theta$ through its pulse area. The
preparation phase is matched to the phase difference between the two laser fields,
allowing the excitation amplitudes to be expressed with real and signed coefficients.
Each laser field resonantly drives one initial state to the same final state through HEB excitation.

The hyperfine interaction $V_{\mathrm{HFI}}=\sum_{k=1,2}\mathbf
T^{(k)}\cdot\mathbf N^{(k)}$ couples electronic tensors $\mathbf T^{(k)}$ to
nuclear tensors $\mathbf N^{(k)}$, where $k=1$ and $2$ correspond to the M1 and E2
interactions, respectively.

\begin{figure}[!tbp]
 \includegraphics[width=\columnwidth]{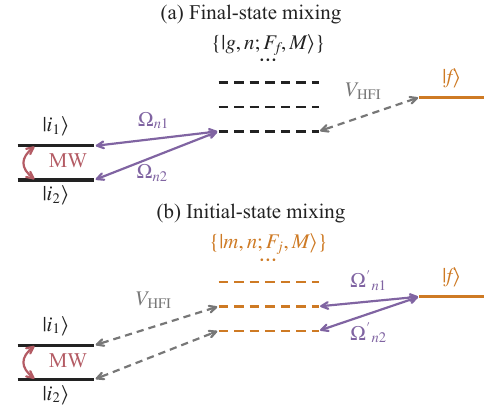}
 \caption{The two time orderings contributing to HEB excitation:
(a) final-state and (b) initial-state hyperfine mixing.
Purple solid arrows denote electronic E1 transitions driven by the two laser fields, indexed by $j=1,2$.
Their Rabi frequencies are $\Omega_{nj}$ in (a) and $\Omega'_{nj}$ in (b).
Gray dashed arrows represent the hyperfine interaction $V_{\mathrm{HFI}}$;
intermediate electronic states are indicated by dashed levels and ellipses.
MW denotes the microwave preparation of the initial superposition.
Orange and black levels, both solid and dashed, correspond to states with the
nucleus in the isomeric and ground states, respectively.
Both time orderings connect the same initial and final states, and their amplitudes are summed coherently.
Level spacings are not to scale.}
 \label{fig:protocol}
\end{figure}

The effective HEB transition moment $D_j$ is
\begin{equation}
 \begin{aligned}
 D_j={}&\sum_n
 \frac{\bra f V_{\mathrm{HFI}}\ket{g,n;F_f,M}
       \bra{g,n;F_f,M}\hat O_0\ket{i_j}}{E_f-E(g,n;F_f)}\\
 &+\sum_n
 \frac{\bra f\hat O_0\ket{m,n;F_j,M}
       \bra{m,n;F_j,M}V_{\mathrm{HFI}}\ket{i_j}}{E_{i_j}-E(m,n;F_j)}.
 \end{aligned}
 \label{eq:moment}
\end{equation}
Here $j=1,2$ labels the two initial states, and $n$ runs over intermediate
electronic states. $\hat O_0$ is the electronic E1 operator for $\pi$-polarized
light.
The first term describes optical absorption followed by the hyperfine interaction,
corresponding to final-state mixing; the second describes the reverse ordering,
corresponding to initial-state mixing [Fig.~\ref{fig:protocol}].
The signed energy mismatch
$\Delta_n=E_f-E(g,n;F_f)$ is set by the atomic and nuclear level energies; for the
two $7p$ states it is approximately $+0.887$ and $-0.702$ eV. These small energy
mismatches enhance the $7p_{1/2,3/2}$ contributions, making final-state mixing the
dominant contribution to the HEB transition moments.
The applied laser frequencies are resonant with the effective
$\ket{i_j}\rightarrow\ket f$ transitions.

Separating the nuclear multipoles in Eq.~\eqref{eq:moment} yields
\begin{equation}
 D_j=C_{j,\mathrm{M1}}N_{\mathrm{M1}}
       +C_{j,\mathrm{E2}}N_{\mathrm{E2}},
 \label{eq:separation}
\end{equation}
where $N_{\mathrm{M1}}$ and $N_{\mathrm{E2}}$ are downward nuclear
transition amplitudes. The electronic coefficients $C_{j,\alpha}$ contain
electronic matrix elements, angular factors, and energy denominators.
We evaluate the coefficients $C_{j,\alpha}$ using experimental energies \cite{Blaise2021},
electronic E1 matrix elements \cite{Li2022}, and hyperfine
matrix elements calculated using the pCI package \cite{Cheung2025PCI},
supplemented by AMBiT calculations \cite{Kahl2019}.
Calculation details are given in Sec. I of the Supplemental Material
\cite{Supplemental}.

\letterheading{Interference and amplitude extraction}For equal laser-field amplitudes
$\mathcal E_1=\mathcal E_2=\mathcal E$, the total effective HEB transition moment
for the superposition in Eq.~\eqref{eq:input} is
\begin{equation}
 \begin{aligned}
 \mathcal D(\theta)&=D_1\sin\theta+D_2\cos\theta\\
 &=N_{\mathrm{M1}}\mathcal C_{\mathrm{M1}}(\theta)
   +N_{\mathrm{E2}}\mathcal C_{\mathrm{E2}}(\theta),
 \end{aligned}
 \label{eq:amplitude}
\end{equation}
with $\mathcal C_{\alpha}(\theta)=C_{1,\alpha}\sin\theta+C_{2,\alpha}\cos\theta$,
where $\alpha=\mathrm{M1},\mathrm{E2}$. The effective Rabi angular frequency is
$\Omega_{\mathrm{eff}}=\mathcal E\mathcal
D/\hbar$.
The two initial states have different relative M1/E2 weights.
Varying $\theta$ therefore controls the relative M1 and E2 contributions to the total excitation amplitude.
A preparation satisfying $\mathcal C_{\mathrm{M1}}=0$ cancels the M1 contribution
while retaining the E2 contribution; similarly, choosing $\mathcal
C_{\mathrm{E2}}=0$ isolates the M1 contribution.
We define the dark angle $\theta_D$ as the mixing angle at which the total
excitation amplitude vanishes, $\mathcal D(\theta_D)=0$. This condition gives
\begin{equation}
 N_{\mathrm{E2}}=-N_{\mathrm{M1}}
       \frac{\mathcal C_{\mathrm{M1}}(\theta_D)}
            {\mathcal C_{\mathrm{E2}}(\theta_D)}.
 \label{eq:inverse}
\end{equation}
The position $\theta_D$ of the interference minimum, together with the calculated
electronic coefficients, determines the signed E2/M1 transition amplitude ratio. An
independent measurement of the M1 transition strength then allows the absolute nuclear E2
transition amplitude to be extracted.
The ratio $N_{\mathrm{E2}}/N_{\mathrm{M1}}$ is invariant under a redefinition of
the phases of the nuclear states.

For numerical comparisons, we take $N_{\mathrm{M1}}>0$ and introduce the
dimensionless nuclear E2 transition amplitude $x=N_{\mathrm{E2}}/N_{\mathrm{E2,ref}}$, where the
positive reference $N_{\mathrm{E2,ref}}$ corresponds to $B_\downarrow(E2)=27$ W.u.
\cite{Minkov2017}. Under our operator convention,
$B_\downarrow(M1)=3|N_{\mathrm{M1}}|^2/[4\pi(2I_m+1)]$ and
$B_\downarrow(E2)=5|N_{\mathrm{E2}}|^2/[4\pi(2I_m+1)]$, with $I_m=3/2$.
The adopted $B_\downarrow(M1)=0.03812\,\mu_N^2$ \cite{Zhang2024,Tiedau2024}
gives $N_{\mathrm{M1}}=\NuclearMOneMuN\,\mu_N$, and
$B_\downarrow(E2)=27x^2$ W.u. The cases $x=\pm1$ therefore have equal
E2 strengths and opposite relative M1/E2 signs.

\begin{figure}[!tbp]
 \includegraphics[width=\columnwidth]{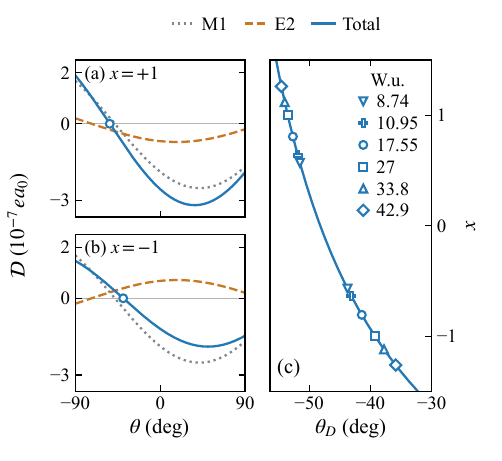}
 \caption{Interference and extraction of the nuclear E2 transition amplitude in \ThThree{}.
(a) and (b) Effective HEB transition moment $\mathcal D$ and its
M1 and E2 contributions plotted against mixing angle at $x=+1$ and $-1$,
respectively. Open circles mark where the total transition moment vanishes.
(c) Dimensionless E2 transition amplitude $x$ as a function of the dark angle
$\theta_D$. Markers denote positive- and negative-sign amplitudes
corresponding to representative E2 strengths
\cite{Chen2025PSM,Zhang2022,Minkov2017,Minkov2021}.
The two laser fields have equal amplitudes throughout.}
 \label{fig:coupling}
 \label{fig:inverse}
\end{figure}

Figures~\ref{fig:coupling}(a) and (b) show how M1 and E2 contributions
interfere as the mixing angle is varied. The individual M1 and E2
contributions vanish at distinct angles, such that cancelling M1 leaves
a finite E2 contribution. At the marked zero of the total excitation
amplitude, the two contributions exactly cancel. At $x=+1$ and $-1$,
the minima occur at $\ThThreeThetaPlus^\circ$ and
$\ThThreeThetaMinus^\circ$, respectively.
Figure~\ref{fig:inverse}(c) shows how the measured dark angle determines
the dimensionless E2 transition amplitude $x$.
A smaller $|dx/d\theta_D|$ reduces the uncertainty in the extracted
nuclear E2 transition amplitude for a given uncertainty in the measured angle.

Let $\mathcal E_j$ and $I_j$ denote the laser-field amplitudes and on-axis
intensities at the ion, respectively. A calibrated ratio
$r=\mathcal E_2/\mathcal E_1=\sqrt{I_2/I_1}$ is included in
Eq.~\eqref{eq:inverse} by replacing
$\mathcal C_\alpha(\theta)$ with
$C_{1,\alpha}\sin\theta+rC_{2,\alpha}\cos\theta$.
A common rescaling of both fields changes the signal strength while preserving the
position of the minimum. Extracting the nuclear E2 transition amplitude therefore
requires calibration of the laser-field amplitude ratio, but not of the absolute
laser intensities.

\letterheading{Excitation signals}To determine how the dark angle can be extracted from a measurable excitation signal, we calculate the final-state population as a function of mixing angle and interrogation time. The resonant dynamics are governed by
\begin{equation}
 H_{\mathrm{eff}}=\frac{\hbar}{2}\sum_{j=1,2}
 \Omega_j\ket f\bra{i_j}+\mathrm{H.c.},\qquad
 \Omega_j=\mathcal E_jD_j/\hbar.
 \label{eq:hamiltonian}
\end{equation}
The initial superposition with coefficients proportional to $(\Omega_1,\Omega_2)$
has the maximal coupling to the final state, with an effective Rabi angular
frequency $\Omega_B=(|\Omega_1|^2+|\Omega_2|^2)^{1/2}$, and is termed the bright
state, denoted by $\ket B$. Its orthogonal superposition has zero coupling.
Under our fixed-phase convention, only the projection of the prepared state onto
$\ket B$ contributes to excitation, with weight $|\langle
B|\psi_i(\theta)\rangle|^2=\sin^2(\theta-\theta_D)$. For equal initial-state loss
rates, the final-state population therefore satisfies
\begin{equation}
 P_f(\theta,t;x)=P_B(t;x)\sin^2[\theta-\theta_D(x)],
 \label{eq:population}
\end{equation}
where $P_B(t;x)$ is the final-state population obtained by exciting the system
from the bright state \cite{Supplemental}.
At a fixed interrogation time, this $\sin^2$ dependence allows $\theta_D$ to be
extracted by fitting data over a broad angular range with appreciable excitation,
avoiding the need to resolve weak signals near the minimum. Alternatively,
sampling on either side of the excitation maximum at $\theta=\theta_B$ could
provide a feedback signal to lock the mixing angle at $\theta_B$, analogous to
spectral-line locking. The dark angle then follows
as $\theta_D=\theta_B-\pi/2$ modulo $\pi$. Substituting the inferred $\theta_D$ into
Eq.~\eqref{eq:inverse} gives the nuclear E2 transition amplitude.

\begin{figure}[!tbp]
 \includegraphics[width=\columnwidth]{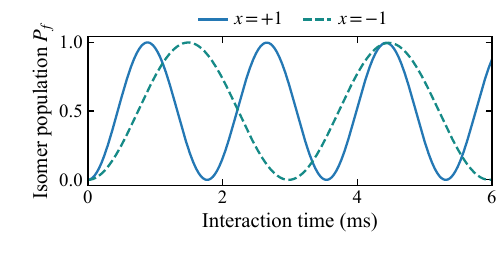}
 \caption{Nuclear isomer population in \ThThree{} including initial-state
spontaneous decay. Solid and dashed curves correspond to $x=+1$ and $-1$.
Each curve begins from its respective bright state.
Each laser has a power of $4$ mW and a $1/e^2$ intensity radius of
$10\,\mu$m.}
 \label{fig:dynamics}
\end{figure}

We therefore evaluate the attainable excitation signal by calculating $P_B(t;x)$
with a master equation that includes decay of each initial state to a loss state.
The adopted $7s_{1/2}$ lifetime is $0.60$ s \cite{Radnaev2012}.
Each laser has a power of $P=4$ mW and a Gaussian intensity profile with a $1/e^2$
radius of $w_0=10\,\mu$m, yielding an on-axis intensity of $I_0=2P/(\pi w_0^2)$.
The fields remain constant during interrogation.
The laser frequencies are tuned to the light-shifted resonances.
Figure~\ref{fig:dynamics} shows appreciable excitation for both signs of $x$.
At $x=+1$, the first maximum reaches
$\ThThreePeakPct\%$ at $\ThThreePeakMs$ ms.

The measurement sequence consists of microwave preparation, simultaneous laser
excitation, and detection of the final-state population. This sequence is repeated
at different mixing angles $\theta$.
The two laser frequencies, corresponding to wavelengths near $225.9$ nm, are
separated by the $17.8$ GHz hyperfine splitting between the two initial states.
These two coherent optical fields are derived from a single laser and delivered to
the ion along a common optical path.
Common-mode power fluctuations do not affect the intensity ratio between the two
beams, and residual fluctuations are suppressed via monitoring and feedback.
Such relative-intensity stabilization has been demonstrated for
electro-optically generated laser pairs \cite{Wang2022Ratio}.
A shared radio-frequency reference maintains phase coherence between the optical
beat note and the microwave preparation field.
For ultraviolet generation, two successive frequency-doubling stages have produced
$0.56$ W of single-frequency continuous-wave light at $229$ nm \cite{Kaneda2016}.

\letterheading{Extraction uncertainties}Uncertainties in the measured dark angle, the calibrated laser-field
amplitude ratio, the calculated electronic matrix elements, and the
relative phase between the excitation pathways propagate through the
inversion relation into uncertainty in the extracted nuclear E2
transition amplitude. We evaluate their contributions separately at
$x=\pm1$. We use $\sigma_y$ to denote the standard uncertainty of a quantity
$y$. The relative standard uncertainty of the extracted E2
transition amplitude is
$\sigma_{N_{\mathrm{E2}}}/|N_{\mathrm{E2}}|=\sigma_x/|x|$,
which reduces to $\sigma_x$ at $x=\pm1$.

\begin{figure}[!hb]
 \includegraphics[width=\columnwidth]{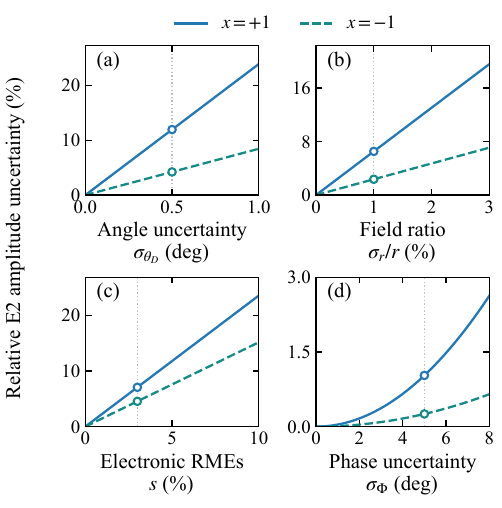}
 \caption{Relative standard uncertainty of the extracted nuclear E2 transition amplitude due to uncertainties in (a) the dark angle, (b) the laser-field
amplitude ratio, (c) electronic reduced matrix elements, and (d) the
relative phase between the excitation pathways. Solid and dashed curves
denote $x=+1$ and $-1$. Vertical dotted lines and open circles mark the
assumed input standard uncertainties $\sigma_{\theta_D}=0.5^\circ$,
$\sigma_r/r=1\%$, $s=3\%$, and
$\sigma_\Phi=\IllustrativePhaseSigmaDeg^\circ$ in (a)--(d), respectively.}
 \label{fig:errors}
\end{figure}

The uncertainty in the measured dark angle propagates as
$\sigma_{x,\theta}=|dx/d\theta_D|\sigma_{\theta_D}$.
For an assumed $\sigma_{\theta_D}=\IllustrativeAngleSigmaDeg^\circ$,
the resulting relative standard uncertainties of the extracted E2 transition amplitude are $\ThThreeAngleSigmaPlusPct\%$ and
$\ThThreeAngleSigmaMinusPct\%$ for $x=+1$ and $-1$, respectively
[Fig.~\ref{fig:errors}(a)].

The calibrated laser-field amplitude ratio contributes
$\sigma_{x,r}=|\partial x/\partial r|\sigma_r$.
We assign a relative standard uncertainty of
$\IllustrativeFieldSigmaPct\%$ to $r$ at equal field amplitudes, $r=1$.
Since $I_2/I_1=r^2$, this corresponds to approximately $2\%$ relative standard
uncertainty in the intensity ratio. Sub-percent power-ratio stability
demonstrated with feedback \cite{Wang2022Ratio} supports the feasibility
of this calibration target. This assumed uncertainty yields relative
standard uncertainties of $\ThThreeFieldSigmaOnePlusPct\%$ and
$\ThThreeFieldSigmaOneMinusPct\%$ in the extracted E2 transition amplitude for $x=+1$ and $-1$, respectively [Fig.~\ref{fig:errors}(b)].
Extracting the E2 transition amplitude therefore requires both
an accurately determined dark angle and a calibrated laser-field amplitude ratio.

The electronic coefficients in Eq.~\eqref{eq:inverse} depend on the electronic
E1 and hyperfine reduced matrix elements in both time orderings.
For each electronic reduced matrix element $R_\ell$, we define its fractional
deviation $\eta_\ell=\delta R_\ell/R_\ell$ and the sensitivity coefficient
$J_\ell=\partial x/\partial\eta_\ell$. We assign independent relative standard
uncertainties of equal size $s$, giving
$\sigma_{x,\mathrm{el}}=s\sqrt{\sum_\ell J_\ell^2}$.
We vary $s$ to assess the extraction sensitivity and adopt
$s=\IllustrativeRMESigmaPct\%$ as a representative value \cite{Supplemental}.
Each electronic matrix element is varied consistently wherever it appears in the uncertainty analysis.
The resulting relative standard uncertainties of the extracted E2 transition amplitude are $\ThThreeSigmaThreePlusPct\%$ and
$\ThThreeSigmaThreeMinusPct\%$ [Fig.~\ref{fig:errors}(c)].

The relative phase $\Phi=\phi_0+\phi_2-\phi_1$ provides the fourth
source of uncertainty. Here $\phi_0$ is the preparation phase and
$\phi_{1,2}$ are the laser phases at the ion location. A nonzero $\Phi$
can raise the excitation minimum and shift its angular position.
We assume a zero-mean Gaussian relative-phase distribution with standard
uncertainty $\sigma_\Phi$. For the representative input
$\sigma_\Phi=\IllustrativePhaseSigmaDeg^\circ$, the resulting relative standard
uncertainties of the extracted E2 transition amplitude are
$\ThThreePhaseSigmaPlusPct\%$ and $\ThThreePhaseSigmaMinusPct\%$ for
$x=+1$ and $-1$, respectively [Fig.~\ref{fig:errors}(d)].
Phase locking with rms phase noise of $2.4^\circ$ has been demonstrated
\cite{Preuschoff2022}. The optical and preparation phases require joint
calibration.

Under the assumed conditions, angle measurements and electronic matrix elements
make the largest contributions to the extraction uncertainty. The uncertainty
propagation and its dependence on $x$ are detailed in Sec. II of the Supplemental
Material \cite{Supplemental}.

\letterheading{Conclusion and outlook}Quantum interference in HEB excitation provides a route to measuring
the nuclear E2 transition amplitude of $^{229}$Th and its sign relative
to the nuclear M1 transition amplitude. Scanning the mixing angle locates the interference
minimum; the E2 transition amplitude is then extracted using
the calculated electronic coefficients, the calibrated laser-field amplitude
ratio, and an independent M1 reference.
In \ThThree{}, our dynamical calculations predict appreciable nuclear
excitation under experimentally accessible laser conditions. Accurate
determination of the mixing angle is central to the extraction precision.
The resulting measurement of the E2 transition amplitude would test nuclear models
describing deformation, collective motion, and configuration mixing
associated with the clock transition.

The approach can be extended to other $^{229}$Th ions supporting
coherent HEB pathways with distinct relative M1 and E2 weights. Their
rich electronic level structures can offer a wider choice of transitions
and opportunities to suppress systematic uncertainties. Strong
electronic-bridge excitation has been predicted in Th$^{35+}$
\cite{Bilous2020}. Its large electronic angular momenta provide numerous
hyperfine and magnetic sublevels, enabling a broad choice of excitation
pathways for realizing the interference scheme. Lower charge states
offer further possibilities for state preparation and readout. The
long-lived $J=0$ metastable states $6d^2\,{}^3P_0$ in Th$^{2+}$
\cite{Maurya2026} and $6p^4\,{}^3P_0$ in Th$^{6+}$ \cite{Yu2025} provide
a single hyperfine level for the initial state, $F=I_g$.
Two coherent laser fields could instead drive HEB excitation from a
$J=0$ state to two final states of the nuclear isomer. The superposition of
these final states would encode the relative excitation amplitudes.
A microwave or Raman analysis pulse, followed by population detection,
could measure the final-state mixing angle. This reverses the configuration of the present scheme,
shifting the angular readout from initial-state preparation to final-state
analysis.

\begin{samepage}
\letterheading{Acknowledgments}We acknowledge support from the Strategic
Priority Research Program of the Chinese Academy of Sciences
(Grant No. XDB0920000), the National Natural Science Foundation of China
(Grant No. 12393821), and the Innovation Fund for Scientific and
Technological Personnel of Hainan Province (Grant No. 225RC636).\par
\end{samepage}

\letterheading{Data availability}The data that support the findings of this
article are not publicly available. The data are available from the authors
upon reasonable request.

\clearpage
\onecolumngrid
\setcounter{equation}{0}
\setcounter{table}{0}
\setcounter{figure}{0}
\setcounter{section}{0}
\setcounter{subsection}{0}
\renewcommand{\theequation}{S\arabic{equation}}
\renewcommand{\thetable}{S\arabic{table}}
\renewcommand{\thefigure}{S\arabic{figure}}
\renewcommand{\theHequation}{supp.\arabic{equation}}
\renewcommand{\theHtable}{supp.\arabic{table}}
\renewcommand{\theHfigure}{supp.\arabic{figure}}
\renewcommand{\theHsection}{supp.\arabic{section}}
\renewcommand{\theHsubsection}{supp.\arabic{section}.\arabic{subsection}}
\setcounter{secnumdepth}{2}
\renewcommand{\thesection}{\Roman{section}}
\begin{center}
\large\bfseries Supplemental Material\\[6pt]
\paperTitle
\end{center}

\begin{center}
\begin{minipage}{400pt}
\small
\setlength{\parindent}{1em}
\indent
We specify the angular-momentum conventions and the two time orderings of the
hyperfine-electronic-bridge (HEB) excitation amplitude, along with the
nuclear-tensor normalization and atomic inputs for \ThThree{}. We describe the
preparation of a coherent superposition and the extraction of the nuclear E2
transition amplitude from an excitation minimum. We also describe the excitation
dynamics including initial-state decay and the propagation of uncertainties in the
measured dark angle, laser-field amplitude ratio, electronic matrix elements, and
relative phase.
\end{minipage}
\end{center}

\section{Calculation details}
\label{sm:calculation-details}

\subsection{Hyperfine-resolved amplitudes}
\label{sm:amplitude}

The nuclear ground and isomeric states are labeled $g$ and $m$, with $I_g=5/2$ and
$I_m=3/2$. We denote a coupled nuclear--electronic state by
$|\nu,e;F,M\rangle=|(I_\nu J_e)F,M\rangle$, where $\nu=g,m$ labels the nuclear state
and $e$ labels the electronic state. The index $n$ runs over intermediate electronic
states. We use the Wigner--Eckart convention
\begin{equation}
 \bra{F,M}O^{(L)}_q\ket{F',M'}=(-1)^{F-M}
 \begin{pmatrix}F&L&F'\\-M&q&M'\end{pmatrix}
 \langle F\Vert O^{(L)}\Vert F'\rangle .
 \label{sm:we}
\end{equation}
The hyperfine interaction (HFI) connecting the two nuclear states is
\begin{equation}
 V_{\mathrm{HFI}}=\sum_{k=1,2}\sum_q(-1)^q T^{(k)}_q N^{(k)}_{-q}.
 \label{sm:hfi}
\end{equation}
The ranks $k=1$ and $k=2$ correspond to nuclear M1 and E2 couplings, respectively;
the absorbed laser photon drives an electronic E1 transition. The scalar hyperfine
matrix element is
\begin{align}
 &\bra{\nu,e;F,M}V^{(k)}_{\mathrm{HFI}}\ket{\nu',e';F,M}\nonumber\\
 &\quad=(-1)^{I_{\nu'}+J_e+F}
 \begin{Bmatrix}I_\nu&J_e&F\\J_{e'}&I_{\nu'}&k\end{Bmatrix}
 \langle I_\nu\Vert N^{(k)}\Vert I_{\nu'}\rangle
 \langle J_e\Vert T^{(k)}\Vert J_{e'}\rangle .
 \label{sm:hfi-rme}
\end{align}
The electronic E1 operator is $O^{(1)}=D^{(1)}$. It leaves the nuclear state unchanged, with
\begin{align}
 &\langle(IJ_e)F\Vert O^{(L)}\Vert(IJ_{e'})F'\rangle\nonumber\\
 &\quad=(-1)^{I+J_{e'}+F+L}\sqrt{(2F+1)(2F'+1)}
 \begin{Bmatrix}J_e&F&I\\F'&J_{e'}&L\end{Bmatrix}
 \langle J_e\Vert O^{(L)}\Vert J_{e'}\rangle .
 \label{sm:optical-rme}
\end{align}
Eqs.~\eqref{sm:we}--\eqref{sm:optical-rme} retain the magnetic-sublevel
and hyperfine interference factors explicitly.

Define $|i_j\rangle=|g,7s_{1/2};F_j,M\rangle$ and
$|f\rangle=|m,5f_{5/2};F_f,M\rangle$ for \ThThree{}.
The two initial states have $(F_1,M)=(2,2)$ and $(F_2,M)=(3,2)$; the final state has
$(F_f,M)=(3,2)$, and the laser fields are $\pi$-polarized. At overall optical resonance,
first-order hyperfine mixing of the initial and final states gives the retained
excitation amplitude
\begin{align}
 D_j={}&\sum_{n\in\mathcal P}\sum_{k=1,2}
 \frac{\bra{f}V^{(k)}_{\mathrm{HFI}}\ket{g,n;F_f,M}
       \bra{g,n;F_f,M}O_0\ket{i_j}}{\Delta_n}\nonumber\\
 &+\sum_{n\in\mathcal Q}\sum_{k=1,2}
 \frac{\bra{f}O_0\ket{m,n;F_j,M}
       \bra{m,n;F_j,M}V^{(k)}_{\mathrm{HFI}}\ket{i_j}}
      {\bar\Delta_{nj}},
 \label{sm:near}
\end{align}
The two terms in Eq.~\eqref{sm:near} represent the two time orderings of the
optical and HFI interactions. In the first, optical absorption is followed by HFI
coupling from the nuclear ground state $g$ to the isomeric state $m$; this is the
final-state-mixing contribution. In the second, HFI coupling precedes optical
absorption, giving the initial-state-mixing contribution. Their amplitudes are
summed coherently.

Here $\Delta_n=E_f-E(g,n;F_f)$ and $\bar\Delta_{nj}=E_{i_j}-E(m,n;F_j)$ are the
energy mismatches between the external and virtual intermediate states. These
mismatches are set by the nuclear transition energy, electronic level energies, and
hyperfine shifts. We retain the intermediate-state sets $\mathcal
P=\{7p_{1/2,3/2},\ldots,10p_{1/2,3/2}\}$ and
$\mathcal Q=\{6d_{3/2,5/2},\ldots,9d_{3/2,5/2}\}$. The second ordering uses
optical angular factors in the $I_m=3/2$ nuclear subspace and conserves $F_j$ and
$M$ during hyperfine mixing. Diagonal hyperfine shifts are included for the
external states and the leading $7p_j$ and $6d_j$ intermediate states.

\subsection{Nuclear tensors and reference inputs}
\label{sm:nuclear}

Downward radiative strengths are defined as
\begin{equation}
 B_\downarrow(\pi k)=
 \frac{|\langle I_g\Vert\mathcal M(\pi k)\Vert I_m\rangle|^2}{2I_m+1},
 \qquad B_\uparrow(\pi k)=\frac23 B_\downarrow(\pi k).
 \label{sm:strength}
\end{equation}
In the convention used here, the nuclear operators are the magnetic moment
$\mu^{(1)}$ and $Q^C_{2q}=\sum_p e r_p^2
C_{2q}(\hat r_p)$, where $C_{kq}=\sqrt{4\pi/(2k+1)}Y_{kq}$. Consequently, $\mathcal
M(M1)=\sqrt{3/(4\pi)}\mu^{(1)}$ and $Q^C_2=\sqrt{4\pi/5}\mathcal M(E2)$. We use the
downward nuclear transition amplitudes $N_k=\langle I_g\Vert N^{(k)}\Vert
I_m\rangle$, so that
\begin{equation}
 \langle I_m\Vert N^{(k)}\Vert I_g\rangle
 =(-1)^{I_m-I_g}N_k^*=-N_k^* .
 \label{sm:nuclear-reverse}
\end{equation}
We denote the rank-one and rank-two nuclear transition amplitudes by $N_{\mathrm{M1}}$ and
$N_{\mathrm{E2}}$, respectively. This relation is applied consistently to both
ranks.

Defining $b_1=B_\downarrow(M1)/\mu_N^2$ and $b_2=B_\downarrow(E2)/(e^2\,\mathrm{fm}^4)$, the conversion relations in the atomic-unit conventions used for the electronic HFI operators are
\begin{align}
 N_{\mathrm{M1}}&=\mu_N
       \sqrt{\frac{4\pi}{3}(2I_m+1)b_1},\nonumber\\
 \frac{N_{\mathrm{E2}}}{ea_0^2}&=s_2\sqrt{\frac{4\pi}{5}(2I_m+1)b_2}
       \left(\frac{1\,\mathrm{fm}}{a_0}\right)^2,
 \qquad s_2=\pm1.
 \label{sm:conversion}
\end{align}
Here the nuclear magneton $\mu_N$ has numerical value $1.98713\times10^{-6}$ in the magnetic
atomic-unit convention used for $T^{(1)}$, and $a_0=5.29177\times10^{-11}$ m. The
corresponding Weisskopf unit for $A=229$ is
\begin{equation}
 B_W(E2)=\frac{1}{4\pi}\left(\frac35\right)^2
 (1.2 A^{1/3}\,\mathrm{fm})^4e^2
 =83.23\,e^2\,\mathrm{fm}^4 .
 \label{sm:weisskopf}
\end{equation}

We use $\nu_N=2.0204074\times10^{15}$ Hz \cite{Zhang2024},
$B_\downarrow(M1)=0.03812\,\mu_N^2$, and the reference strength
$B_\downarrow(E2)=27$ W.u. \cite{Minkov2017,Bilous2018}. The M1 reference is
obtained from the crystal radiative lifetime of $641$ s \cite{Zhang2024}, using the
adopted refractive index $n=1.586$ and the conversion
$\tau_{\mathrm{vac}}=n^3\tau_{\mathrm{crystal}}$ for M1-dominated decay
\cite{Tiedau2024}. With these reference inputs, the standard free-space multipole
decay rates \cite{Bilous2018} give $\Gamma(M1)\simeq3.9\times10^{-4}$ s$^{-1}$ and
$\Gamma(E2)\simeq1.1\times10^{-13}$ s$^{-1}$, hence
$\Gamma(M1)/\Gamma(E2)\simeq3.5\times10^9$. Equation~\eqref{sm:conversion} gives
$N_{\mathrm{M1}}=1.588\times10^{-6}$ in the magnetic atomic units defined above,
equivalent to $0.7992\,\mu_N$, and the positive reference
$N_{\mathrm{E2,ref}}=5.367\times10^{-8}\,ea_0^2$. We parameterize the nuclear E2 transition amplitude
by $x=N_{\mathrm{E2}}/N_{\mathrm{E2,ref}}$ at fixed $N_{\mathrm{M1}}$; thus
$B_\downarrow(E2)=27x^2$ W.u. The relative sign is defined using the nuclear-tensor
and electronic-state phase conventions of Eqs.~\eqref{sm:hfi-rme} and
\eqref{sm:optical-rme}.

For diagonal hyperfine shifts, we use $\mu_g=0.366\mu_N$, $Q_g=3.11\,e\,\mathrm b$
\cite{Porsev2021}, $\mu_m=-0.378\mu_N$ \cite{Yamaguchi2024}, and
$Q_m=1.773\,e\,\mathrm b$, obtained from $Q_m/Q_g=0.57003$ \cite{Zhang2024}. Here
$\mathrm b$ denotes the barn.

\subsection{Atomic inputs}
\label{sm:atomic}

We treat \ThThree{} as a single valence electron outside a closed-shell radon-like
core. To calculate the leading electronic HFI matrix elements, we use the pCI package with
the linearized coupled-cluster singles-and-doubles (LCCSD) method
\cite{Cheung2025PCI}. We also use AMBiT with configuration interaction and
many-body perturbation theory to calculate electronic E1 and HFI matrix elements
\cite{Kahl2019}. Correlation effects beyond the frozen-core
Dirac--Hartree--Fock approximation are incorporated through the one-body operator
$\Sigma_1$, giving the effective valence Hamiltonian
\begin{equation*}
 H_{\mathrm{eff}}=H_{\mathrm{DHF}}+\Sigma_1.
\end{equation*}
In the pCI calculation, $H_{\mathrm{DHF}}$ includes the Breit interaction. The LCCSD
amplitudes describe single and double excitations of core and valence electrons and
sum the associated correlation contributions to all orders. We use the resulting
$\Sigma_1$ matrix elements for the selected external valence orbitals specified
below; the remaining matrix elements are evaluated using second-order many-body
perturbation theory.
For the pCI calculation, the radial basis contains 50 B-splines of order 8 in a $45a_0$
cavity. The internal correlation sums extend to $n=35$, $\ell=6$, and the
configuration-interaction basis is $25spdf$. The LCCSD treatment of $\Sigma_1$ uses
the external valence orbitals $7s$--$9s$ and $6d$--$7d$ in the even-parity
sector, and $7p$--$9p$, $5f_{5/2}$, $5f_{7/2}$, and $6f_{5/2}$ in the odd-parity
sector, including both fine-structure partners of each $p$ and $d$ orbital.

In the AMBiT calculation, $\Sigma_1$ is evaluated using second-order many-body
perturbation theory. We scale this one-body correlation potential to reproduce the
experimental energy of the lowest state in each $\kappa$ symmetry
\cite{YuClock2024,YuBridge2026}. Diagonalizing the effective Hamiltonian gives the
correlated valence states used to evaluate the electronic matrix elements.
The AMBiT calculation includes the Breit interaction and adopts a $20spdf$
valence basis together with a $30spdfghi$
perturbative basis. Its radial basis consists of 80 B-splines of order 11 in a
$120a_0$ cavity.

The pCI HFI operators and both the AMBiT E1 and HFI operators include static
random-phase-approximation (RPA) corrections, accounting for the self-consistent
response of the core electrons to the electric or hyperfine field. The adopted
matrix elements and spectroscopic energies are specified below.

Table~\ref{sm:th3-levels} lists all electronic energies and diagonal hyperfine
structure (HFS) constants used in the calculation. Experimental energies are
adopted where available \cite{Blaise2021,Li2022}; the $9p_j$, $10p_j$, and $9d_j$
energies are calculated with AMBiT and referenced to its $5f_{5/2}$ ground state.
For the higher intermediate states, diagonal hyperfine splittings are neglected
compared with their electronic energy mismatches.
The HFS constants for $5f_{5/2}$ and $6d_j$ are
experimental values \cite{Campbell2011}. For $7s_{1/2}$, we adopt the recommended
value $A/g=40563$ MHz from relativistic coupled-cluster calculations including
single, double, and triple excitations (CCSDT) \cite{Porsev2026}. Multiplying by
$g=\mu_g/(I_g\mu_N)=0.1464$ gives $A=5938$ MHz after rounding. The $7p_j$ magnetic
hyperfine constants in Ref.~\cite{Safronova2013} were originally calculated using
$\mu_g=0.360\,\mu_N$. Since these constants scale linearly with $\mu_g$, we rescale
them by $0.366/0.360$ to match the updated magnetic moment adopted here
\cite{Porsev2021}. The $7p_{3/2}$ quadrupole constant already uses
$Q_g=3.11\,e\,\mathrm b$.

\begin{table}[htb]
\caption{Electronic centroids and ground-nuclear-sector HFS constants for
\ThThree{}. Energies are given relative to $5f_{5/2}$. The $B$ entries for $J=1/2$
vanish by angular-momentum selection rules. Dashes indicate neglected diagonal
HFS splittings for the higher intermediate states.}
\label{sm:th3-levels}
\begin{ruledtabular}
\begin{tabular}{lrrrl}
State & $E/(hc)$ (cm$^{-1}$) & $A$ (MHz) & $B$ (MHz) & Energy source \\
\hline
$5f_{5/2}$ & 0 & 82.2 & 2269 & Expt.~\cite{Li2022} \\
$6d_{3/2}$ & 9193.245 & 155.3 & 2265 & Expt.~\cite{Li2022} \\
$6d_{5/2}$ & 14486.34 & -12.6 & 2694 & Expt.~\cite{Li2022} \\
$7s_{1/2}$ & 23130.75 & 5938 & 0 & Expt.~\cite{Li2022} \\
$7p_{1/2}$ & 60239.1 & 1443 & 0 & Expt.~\cite{Li2022} \\
$7p_{3/2}$ & 73055.9 & 121.4 & 5310 & Expt.~\cite{Li2022} \\
$7d_{3/2}$ & 119684.6 & --- & --- & Expt.~\cite{Li2022} \\
$7d_{5/2}$ & 121427.1 & --- & --- & Expt.~\cite{Li2022} \\
$8p_{1/2}$ & 134516.5 & --- & 0 & Expt.~\cite{Li2022} \\
$8p_{3/2}$ & 139870.9 & --- & --- & Expt.~\cite{Li2022} \\
$9p_{1/2}$ & 157154 & --- & 0 & AMBiT (this work) \\
$9p_{3/2}$ & 159871 & --- & --- & AMBiT (this work) \\
$8d_{3/2}$ & 161954 & --- & --- & Expt.~\cite{Li2022} \\
$8d_{5/2}$ & 162527.8 & --- & --- & Expt.~\cite{Li2022} \\
$9d_{3/2}$ & 171649 & --- & --- & AMBiT (this work) \\
$9d_{5/2}$ & 172123 & --- & --- & AMBiT (this work) \\
$10p_{1/2}$ & 175543 & --- & 0 & AMBiT (this work) \\
$10p_{3/2}$ & 177156 & --- & --- & AMBiT (this work)

\end{tabular}
\end{ruledtabular}
\end{table}

The diagonal hyperfine shift expressed in frequency units is
\begin{equation}
 \frac{E_{\mathrm{HFS}}(I,J,F)}{h}
 =\frac{A K}{2}
 +B\frac{\tfrac34 K(K+1)-I(I+1)J(J+1)}
 {2I(2I-1)J(2J-1)},
 \quad K=F(F+1)-I(I+1)-J(J+1).
 \label{sm:AB}
\end{equation}
The second term is omitted for $J=1/2$. For the $5f_{5/2}$ final
state and $6d_j$ virtual intermediate states with the nucleus in the isomeric state,
the hyperfine constants are scaled as
$A_m=A_g(\mu_m/I_m)/(\mu_g/I_g)$ and $B_m=B_gQ_m/Q_g$. The same electronic centroids
are used for the two nuclear sectors in the energy denominators; the nuclear
transition energy is added in the nuclear isomer sector.

Fourteen electronic E1 reduced matrix-element magnitudes are taken from the
coupled-cluster singles-and-doubles (CCSD) values in Table III of
Ref.~\cite{Li2022}; the two $10p_j$ matrix elements are calculated with AMBiT.

The signs of all electronic E1 and HFI matrix elements follow a consistent AMBiT
state-phase convention \cite{Kahl2019}. The six nonzero HFI magnitudes involving
$7p_j$ and $6d_j$ are taken from pCI; the remaining 18 HFI magnitudes are obtained
with AMBiT.

Tables~\ref{sm:th3-rme} and \ref{sm:th3-order2-rme} give the resulting signed inputs
alongside the AMBiT HFI values for comparison.

\begin{table}[htb]
\caption{Signed electronic inputs for optical absorption followed by HFI coupling
between the nuclear ground and isomeric states in \ThThree{}. $D_n=\langle n\Vert
D\Vert7s_{1/2}\rangle$ is given in $ea_0$; $T_{kn}=\langle5f_{5/2}\Vert
T^{(k)}\Vert n\rangle$ uses the AMBiT hyperfine atomic-unit convention consistent
with Eq.~\eqref{sm:conversion}.}
\label{sm:th3-rme}
\begin{ruledtabular}
\begin{tabular}{lrrrrr}
 & & \multicolumn{2}{c}{Adopted} & \multicolumn{2}{c}{AMBiT} \\
Intermediate state & $D_n$ & $T_{1n}$ & $T_{2n}$ & $T_{1n}$ & $T_{2n}$ \\
\hline
$7p_{1/2}$ & $-2.438$ & $0$ & $+0.114$ & $0$ & $+0.136$ \\
$7p_{3/2}$ & $-3.392$ & $-0.0339$ & $+0.335$ & $-0.0322$ & $+0.317$ \\
$8p_{1/2}$ & $-0.2206$ & $0$ & $-0.0775$ & $0$ & $-0.0775$ \\
$8p_{3/2}$ & $+0.08750$ & $+0.0224$ & $-0.198$ & $+0.0224$ & $-0.198$ \\
$9p_{1/2}$ & $-0.1689$ & $0$ & $-0.0475$ & $0$ & $-0.0475$ \\
$9p_{3/2}$ & $-0.04190$ & $+0.0165$ & $-0.139$ & $+0.0165$ & $-0.139$ \\
$10p_{1/2}$ & $-0.1024$ & $0$ & $-0.0324$ & $0$ & $-0.0324$ \\
$10p_{3/2}$ & $-0.02407$ & $+0.0127$ & $-0.105$ & $+0.0127$ & $-0.105$

\end{tabular}
\end{ruledtabular}
\end{table}

\begin{table}[htb]
\caption{Signed electronic inputs for HFI coupling between the nuclear ground and
isomeric states followed by optical absorption in \ThThree{}. $\bar
D_n=\langle5f_{5/2}\Vert D\Vert n\rangle$ is given in $ea_0$; $\bar
T_{kn}=\langle n\Vert T^{(k)}\Vert7s_{1/2}\rangle$ uses the same hyperfine units
as Table~\ref{sm:th3-rme}. The $k=1$ elements for the $d_{5/2}$ states vanish by the
electronic triangle rule.}
\label{sm:th3-order2-rme}
\begin{ruledtabular}
\begin{tabular}{lrrrrr}
 & & \multicolumn{2}{c}{Adopted} & \multicolumn{2}{c}{AMBiT} \\
Intermediate state & $\bar D_n$ & $\bar T_{1n}$ & $\bar T_{2n}$ & $\bar T_{1n}$ & $\bar T_{2n}$ \\
\hline
$6d_{3/2}$ & $-1.562$ & $-0.167$ & $-0.561$ & $-0.158$ & $-0.359$ \\
$6d_{5/2}$ & $-0.4198$ & $0$ & $+3.60$ & $0$ & $+3.41$ \\
$7d_{3/2}$ & $-0.2179$ & $+0.0561$ & $-0.442$ & $+0.0561$ & $-0.442$ \\
$7d_{5/2}$ & $-0.04260$ & $0$ & $-2.53$ & $0$ & $-2.53$ \\
$8d_{3/2}$ & $-0.1525$ & $+0.0344$ & $-0.245$ & $+0.0344$ & $-0.245$ \\
$8d_{5/2}$ & $-0.03170$ & $0$ & $-1.70$ & $0$ & $-1.70$ \\
$9d_{3/2}$ & $-0.1313$ & $+0.0243$ & $-0.153$ & $+0.0243$ & $-0.153$ \\
$9d_{5/2}$ & $-0.02940$ & $0$ & $-1.25$ & $0$ & $-1.25$

\end{tabular}
\end{ruledtabular}
\end{table}

The pCI hyperfine output gives reduced matrix elements $H_1$ and $H_2$ in MHz, with
reference nuclear factors included. The six adopted pCI electronic magnitudes in
Tables~\ref{sm:th3-rme} and \ref{sm:th3-order2-rme} are obtained via
\begin{equation}
 |T_1|=\frac{|H_1|}{\mu_N(E_h/h)_{\mathrm{MHz}}},
 \qquad
 |T_2|=\frac{|H_2|}{(\mathrm b/a_0^2)(E_h/h)_{\mathrm{MHz}}},
 \label{sm:recover}
\end{equation}
where $(E_h/h)_{\mathrm{MHz}}\simeq6.57968\times10^9$ and $\mathrm
b/a_0^2\simeq3.57106\times10^{-8}$. The rank-two tensor uses the normalized
spherical harmonic $C_{2q}$, consistent with Eq.~\eqref{sm:conversion}. For
Hermitian spherical tensors, we use the relation $\langle J_b\Vert O^{(k)}\Vert
J_c\rangle=(-1)^{J_b-J_c}\langle J_c\Vert O^{(k)}\Vert J_b\rangle^*$.

The energy mismatches are approximately $\Delta_{7p_{1/2}}=+0.887$ eV and
$\Delta_{7p_{3/2}}=-0.702$ eV. The electronic triangle rule forbids $\langle5f_{5/2}\Vert
T^{(1)}\Vert7p_{1/2}\rangle$, whereas both tensor ranks are allowed through the
$7p_{3/2}$ intermediate state.

Comparisons between LCCSD and higher-order coupled-cluster results for selected
\ThThree{} hyperfine constants show average relative differences below 3\%
\cite{Porsev2021,Porsev2026}. For example, the LCCSD and CCSDT values of $B/Q$ for
the four $5f_j$ and $6d_j$ states reported in Ref.~\cite{Porsev2021} differ by 1.7\%
on average. These comparisons motivate the uncertainty scale for the leading
pCI HFI inputs. The higher intermediate states, whose HFI matrix elements are
calculated with AMBiT, contribute only small corrections. Relative to the sums containing only $7p_j$ and $6d_j$,
the higher intermediate states through $10p_j$ and $9d_j$ contribute 0.39\% to each
M1 amplitude and 3.0\% and 1.0\% to the E2 amplitudes of the two pathways,
respectively.

For the complete input set, we adopt a uniform relative standard
uncertainty of 3\% as a representative setting for the sensitivity analysis and
vary this assumed uncertainty to quantify its effect on the
extracted nuclear E2 transition amplitude [Fig. 5(c) of the main text]. The
uncertainty propagation is specified in Sec.~\ref{sm:uncertainty}.

\subsection{Mixing angle, signed inference, and field calibration}
\label{sm:inference}

Eq.~\eqref{sm:near} is linear in the nuclear tensors:
$D_j(x)=D_j^{\mathrm{M1}}+xD_j^{\mathrm{E2,ref}}$, where $D_j^{\mathrm{M1}}$
contains $N_{\mathrm{M1}}$ and $D_j^{\mathrm{E2,ref}}$ contains
$N_{\mathrm{E2,ref}}$. These effective HEB transition moments are given in $ea_0$.
Before absorbing the preparation and optical phases into the basis states, the general prepared state is
\begin{equation}
 \ket{\psi_i(\theta)}=\sin\theta\ket{i_1}
   +e^{i\phi_0}\cos\theta\ket{i_2}.
 \label{sm:preparation}
\end{equation}
Writing the optical couplings as
$\widetilde\Omega_j=\Omega_j e^{i\phi_j}$ with real, signed
$\Omega_j$, the excitation amplitude is proportional to
\begin{equation}
 e^{i\phi_1}\left[\Omega_1\sin\theta
 +e^{i\Phi}\Omega_2\cos\theta\right],\qquad
 \Phi=\phi_0+\phi_2-\phi_1.
\end{equation}
We set $\Phi=0$ and absorb the matched phases into the basis-state convention. This
gives the real superposition used in the main text and the real coefficients of
Eq.~\eqref{sm:rabi}. Negative coefficients retain their signs and encode a relative
phase of $\pi$.

For a field-amplitude ratio
$r=\mathcal E_2/\mathcal E_1=\sqrt{I_2/I_1}$, where $I_j$ is the intensity of laser
field $j$ at the ion, the Rabi angular frequencies can be written as
\begin{equation}
 \Omega_1=\frac{\mathcal E_1}{\hbar}(D_1^{\mathrm{M1}}+xD_1^{\mathrm{E2,ref}}),\qquad
 \Omega_2=\frac{r\mathcal E_1}{\hbar}(D_2^{\mathrm{M1}}+xD_2^{\mathrm{E2,ref}}).
 \label{sm:rabi}
\end{equation}
The mixing angle $\theta_D$ at which the total excitation amplitude vanishes satisfies
\begin{equation}
 (D_1^{\mathrm{M1}}+xD_1^{\mathrm{E2,ref}})\sin\theta_D+r(D_2^{\mathrm{M1}}+xD_2^{\mathrm{E2,ref}})\cos\theta_D=0,
 \label{sm:dark}
\end{equation}
which gives the inversion relation
\begin{equation}
 x=-\frac{D_1^{\mathrm{M1}}\sin\theta_D+rD_2^{\mathrm{M1}}\cos\theta_D}
          {D_1^{\mathrm{E2,ref}}\sin\theta_D+rD_2^{\mathrm{E2,ref}}\cos\theta_D}.
 \label{sm:inverse}
\end{equation}
The vectors $\mathbf D^{\mathrm{M1}}=(D_1^{\mathrm{M1}},D_2^{\mathrm{M1}})$ and
$\mathbf D^{\mathrm{E2,ref}}=(D_1^{\mathrm{E2,ref}},D_2^{\mathrm{E2,ref}})$ contain
the contributions from the two pathways. Their nonparallel directions make the
dark angle depend on $x$. A continuous branch of $\theta_D$, defined modulo
$\pi$, is used for the inversion. Figure 3(c) of the main text shows $x$ as a
function of the measured dark angle. A common scaling of both optical fields changes
the excitation timescale while leaving Eq.~\eqref{sm:dark} unchanged.

\subsection{Resonant excitation dynamics}
\label{sm:dynamics}

In the rotating frame of the two resonant laser fields,
\begin{equation}
 H/\hbar=\frac12\left(\Omega_1\ket f\bra{i_1}
          +\Omega_2\ket f\bra{i_2}+\mathrm{H.c.}\right).
 \label{sm:hamiltonian}
\end{equation}
For real couplings, the coupled superposition is
$|B\rangle=(\Omega_1|i_1\rangle+\Omega_2|i_2\rangle)/\Omega_B$, with
$\Omega_B=\sqrt{\Omega_1^2+\Omega_2^2}$. The orthogonal superposition has zero
coherent coupling to $|f\rangle$. Decay is included through independent loss jump
operators $L_j=\sqrt{\gamma_j}|\ell_j\rangle\langle i_j|$ and
$L_f=\sqrt{\gamma_f}|\ell_f\rangle\langle f|$, with
\begin{equation}
 \dot\rho=-\frac{i}{\hbar}[H,\rho]
 +\sum_\alpha\left(L_\alpha\rho L_\alpha^\dagger
 -\frac12\{L_\alpha^\dagger L_\alpha,\rho\}\right).
 \label{sm:lindblad}
\end{equation}
The reported signal is $P_f(t)=\langle f|\rho(t)|f\rangle$ for an
initially prepared ion; it is not renormalized to the surviving population.

We use $\gamma_1=\gamma_2=(0.60\,\mathrm s)^{-1}$, based on the $7s$ electronic
lifetime measured in $^{232}$Th$^{3+}$ \cite{Radnaev2012}. Decay of the final state
is negligible over the interrogation interval, so we set $\gamma_f=0$.

For an initial bright state $|B\rangle$ and equal initial-state loss rates
$\gamma_i$, the final-state population is
\begin{equation}
 P_B(t)=\left|
 -\frac{i\Omega_B}{2}
 e^{-(\gamma_i+\gamma_f)t/4}
 \frac{\sinh(\zeta t)}{\zeta}\right|^2,
 \qquad
 \zeta^2=\frac{(\gamma_i-\gamma_f)^2}{16}
          -\frac{\Omega_B^2}{4}.
 \label{sm:analytic-population}
\end{equation}
This expression covers both oscillatory and overdamped evolution and provides a
check on the density-matrix propagation. It also determines the first maximum of
$P_B(t)$ used as the reference time in the population calculations.

For a general prepared state in Eq.~\eqref{sm:preparation} with matched
phases and equal initial-state loss rates, the population factorizes as
\begin{equation}
 P_f(\theta,t;x)=P_B(t;x)\sin^2[\theta-\theta_D(x)].
 \label{sm:scan-population}
\end{equation}
The measured angular dependence allows $\theta_D$ to be extracted by fitting, with $P_B$ setting the signal scale.

Each laser field has power $P=4$ mW and a Gaussian intensity profile with $1/e^2$
radius $w_0=10\,\mu\mathrm m$. Its on-axis intensity is $I_0=2P/(\pi w_0^2)$ and its
peak electric field is $\mathcal E_0=\sqrt{2I_0/(\epsilon_0c)}$. In
Eq.~\eqref{sm:rabi}, the tabulated transition moments in atomic units are multiplied
by $ea_0$ when using SI field amplitudes. Both laser wavelengths are near $225.9$
nm, with a frequency separation of $17.8$ GHz. Constant resonant drives are applied
throughout the interrogation interval in the dynamics calculation.

\section{Uncertainty analysis}
\label{sm:uncertainty-analysis}

With fixed $N_{\mathrm{M1}}$ and $N_{\mathrm{E2,ref}}$, the relation
$N_{\mathrm{E2}}=xN_{\mathrm{E2,ref}}$ implies
$\sigma_{N_{\mathrm{E2}}}/|N_{\mathrm{E2}}|=\sigma_x/|x|$. At $|x|=1$, this relative
standard uncertainty equals $\sigma_x$ numerically.
We evaluate each uncertainty contribution separately at nominal $r=1$.

\subsection{Angle measurement}
\label{sm:angle-measurement}

For equal field amplitudes,
\begin{equation}
 \frac{d\theta_D}{dx}
 =\frac{D_2^{\mathrm{M1}}D_1^{\mathrm{E2,ref}}-D_1^{\mathrm{M1}}D_2^{\mathrm{E2,ref}}}{(D_1^{\mathrm{M1}}+xD_1^{\mathrm{E2,ref}})^2+(D_2^{\mathrm{M1}}+xD_2^{\mathrm{E2,ref}})^2}.
 \label{sm:slope}
\end{equation}

The angular-fit uncertainty $\sigma_{\theta_D}$ propagates as
\begin{equation}
 \sigma_{x,\theta}=\left|\frac{dx}{d\theta_D}\right|\sigma_{\theta_D} .
 \label{sm:angle-uncertainty}
\end{equation}
Here $\sigma_{\theta_D}$ is expressed in radians when Eq.~\eqref{sm:slope} is used.
Figure 5(a) of the main text illustrates this propagation: a smaller
$|dx/d\theta_D|$ gives a smaller uncertainty in the extracted nuclear E2 transition amplitude for a given angular uncertainty.

\subsection{Laser-field amplitude ratio}
\label{sm:field-ratio}

For Fig. 5(b) of the main text, the calibrated field-amplitude ratio $r=\mathcal
E_2/\mathcal E_1$ has standard uncertainty $\sigma_r$ at $r=1$. At a fixed measured
$\theta_D$, differentiating Eq.~\eqref{sm:inverse} gives
\begin{equation}
 J_r=-\frac{\cos\theta_D(D_2^{\mathrm{M1}}+xD_2^{\mathrm{E2,ref}})}
 {D_1^{\mathrm{E2,ref}}\sin\theta_D+D_2^{\mathrm{E2,ref}}\cos\theta_D},\qquad
 \delta x=J_r\delta r,\qquad \sigma_{x,r}=|J_r|\sigma_r .
 \label{sm:field-uncertainty}
\end{equation}
The relative and absolute uncertainties of $r$ coincide at $r=1$. Since
$I_2/I_1=r^2$, the relative standard uncertainty of the intensity ratio is
approximately $2\sigma_r/r$. Evaluated at the respective dark angles for $x=+1$ and
$-1$, the sensitivity coefficients $|J_r|$ are $\ThThreeFieldSensitivityPlus$ and
$\ThThreeFieldSensitivityMinus$, respectively.

\subsection{Electronic matrix elements}
\label{sm:uncertainty}

Let $R_\ell$ denote each independent nonzero electronic reduced matrix element
entering Eq.~\eqref{sm:near}, and define its fractional deviation $\eta_\ell=\delta
R_\ell/R_\ell$. The measured $\theta_D$ and field-amplitude ratio are held fixed
while the electronic matrix elements in the inversion relation are varied.
Spectroscopic hyperfine parameters and the hyperfine basis also remain fixed. With
$\mathbf w=(\sin\theta_D,r\cos\theta_D)$, differentiation gives
\begin{equation}
 J_\ell\equiv\frac{\partial x}{\partial\eta_\ell}
 =-\frac{\mathbf w\cdot
       (\partial_{\eta_\ell}\mathbf D^{\mathrm{M1}}+x\partial_{\eta_\ell}\mathbf D^{\mathrm{E2,ref}})}
       {\mathbf w\cdot\mathbf D^{\mathrm{E2,ref}}},
 \qquad
 \sigma_{x,\mathrm{el}}^2=\sum_{\ell,\ell'}J_\ell\Sigma_{\ell\ell'}J_{\ell'},
 \label{sm:uncertainty-propagation}
\end{equation}
where $\Sigma_{\ell\ell'}$ is the covariance of the fractional input deviations. We
neglect correlations between distinct electronic matrix elements. With a uniform
relative standard uncertainty $s$, $\Sigma_{\ell\ell'}=s^2\delta_{\ell\ell'}$, and
the general expression reduces to $\sigma_{x,\mathrm{el}}=s\sqrt{\sum_\ell
J_\ell^2}$, as used in the main text and Fig. 5(c). The derivatives are evaluated at
the corresponding $\theta_D(x)$ for each value of $x$, including the two reference
cases $x=+1$ and $-1$. Each matrix element is varied consistently wherever it appears; its reverse
matrix element is obtained from the relation given above. Matrix elements that
vanish by selection rules remain zero. The varied inputs are the electronic E1 and
$T^{(1)}$, $T^{(2)}$ reduced matrix elements in both retained time orderings,
comprising 16 E1 and 24 nonzero HFI matrix elements listed in
Tables~\ref{sm:th3-rme} and \ref{sm:th3-order2-rme}. At $s=3\%$, the relative standard
uncertainties of the extracted nuclear E2 transition amplitude are
$\ThThreeSigmaThreePlusPct\%$ and $\ThThreeSigmaThreeMinusPct\%$ for $x=+1$ and
$-1$, respectively.

\subsection{Relative phase}
\label{sm:phase}

The relative phase $\Phi=\phi_0+\phi_2-\phi_1$ combines the microwave preparation
phase and the two optical phases at the ion location. For a given $\Phi$, real
signed couplings $\Omega_j$, and equal initial-state loss rates, the angular
population profile is
\begin{equation}
 \begin{aligned}
 P_f(\theta,t;x,\Phi)
 ={}&\frac{P_B(t;x)}{\Omega_1^2+\Omega_2^2}
 \bigl[\Omega_1^2\sin^2\theta+\Omega_2^2\cos^2\theta\\
 &\qquad+2\Omega_1\Omega_2\sin\theta\cos\theta\cos\Phi\bigr].
 \end{aligned}
 \label{sm:phase-population}
\end{equation}
A nonzero relative phase can raise the minimum population and change its angular
position. At fixed interrogation time and phase, we fit the angular profile with
$P_f(\theta)=P_{\min}+\mathcal A\sin^2(\theta-\theta_{\min})$, allowing a nonzero
$P_{\min}$. The minimum $\theta_{\min}(\Phi)$ satisfies
\begin{equation}
 (\Omega_1^2-\Omega_2^2)\sin2\theta_{\min}
 +2\Omega_1\Omega_2\cos\Phi\cos2\theta_{\min}=0,
 \label{sm:phase-minimum}
\end{equation}
The minimum is chosen on the branch continuous from
$\theta_{\min}(0)=\theta_D$. Applying the nominal zero-phase
inversion, Eq.~\eqref{sm:inverse}, to $\theta_{\min}(\Phi)$ at $r=1$
gives the inferred amplitude $X(\Phi)$. We treat $\Phi$ as an uncertain parameter that
remains constant during an angular scan, with
$\Phi\sim\mathcal N(0,\sigma_\Phi^2)$. We propagate its uncertainty
to second order in the small phase. For $\Phi$ in radians, the leading
corrections are
\begin{align}
 \theta_{\min}(\Phi)-\theta_D
 &=-\frac{\sin4\theta_D}{8}\Phi^2+O(\Phi^4),\nonumber\\
 X(\Phi)-x
 &=A\Phi^2+O(\Phi^4).
 \label{sm:phase-expansion}
\end{align}
Here $A=-\sin(4\theta_D)/(8\,d\theta_D/dx)$. In the quadratic approximation,
the propagated mean and standard deviation about that mean are
\begin{equation}
 \langle X\rangle-x\simeq A\sigma_\Phi^2,\qquad
 \sigma_{x,\Phi}\simeq\sqrt{2}|A|\sigma_\Phi^2.
 \label{sm:phase-uncertainty}
\end{equation}
Figure 5(d) of the main text shows the resulting relative standard
uncertainty $\sigma_{x,\Phi}/|x|$ for $0\le\sigma_\Phi\le8^\circ$.
We use $\sigma_\Phi=\IllustrativePhaseSigmaDeg^\circ$ as a representative
input standard uncertainty.

\subsection{Dependence on the nuclear E2 transition amplitude}
\label{sm:uncertainty-maps}

\begin{figure}[htbp]
 \centering
 \includegraphics[width=0.82\textwidth]{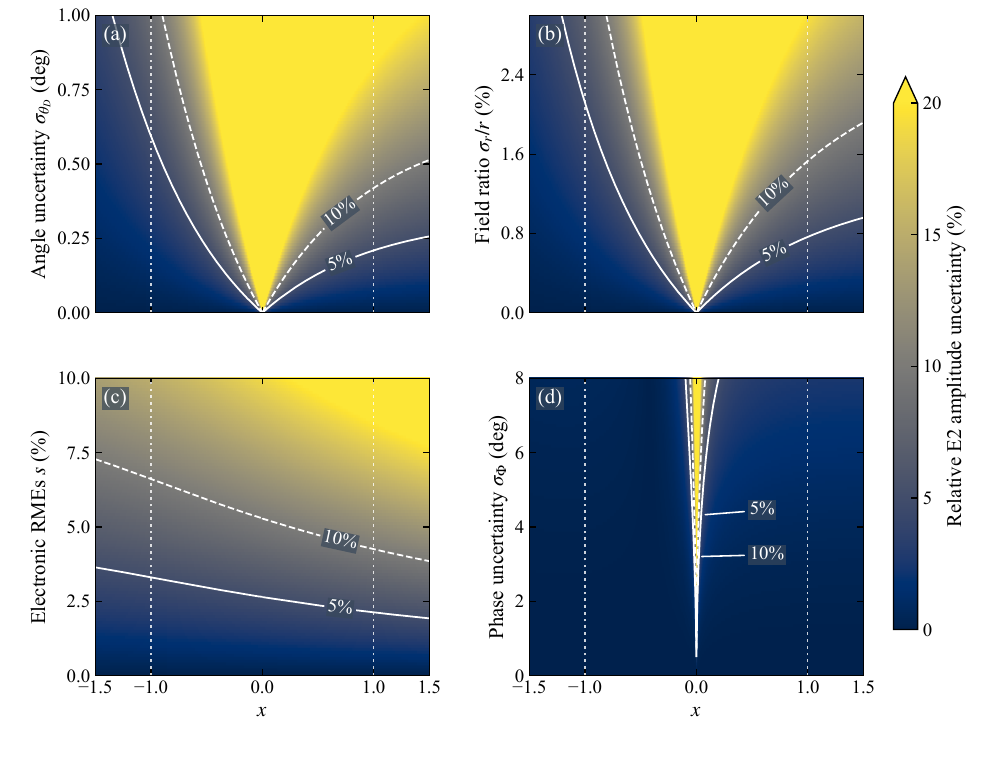}
 \caption{Relative standard uncertainty of the extracted nuclear E2
 transition amplitude as a function of the dimensionless E2 transition amplitude $x$
 and the input standard uncertainty in (a) the dark angle, (b) the
 laser-field amplitude ratio, (c) the electronic reduced matrix elements,
 and (d) the relative phase. Each contribution is evaluated separately
 at nominal $r=1$, with the nuclear reference values and spectroscopic
 inputs fixed. Colors show the corresponding relative standard
 uncertainties $\sigma_{x,\theta}/|x|$, $\sigma_{x,r}/|x|$,
 $\sigma_{x,\mathrm{el}}/|x|$, and $\sigma_{x,\Phi}/|x|$ in (a)--(d),
 respectively. Labeled contours indicate $5\%$ and $10\%$, and vertical
 dashed lines mark $x=\pm1$, corresponding to the cuts shown in Fig. 5
 of the main text. Panel (d) uses second-order small-phase propagation.
 The relative uncertainty is undefined at $x=0$.}
 \label{fig:sm-uncertainty-map}
\end{figure}

Figure~\ref{fig:sm-uncertainty-map} extends the uncertainty analysis beyond
the two representative amplitudes $x=\pm1$. For each candidate value of
$x$, the contours give the input precision required to keep the
corresponding contribution to the relative standard uncertainty of the
extracted nuclear E2 transition amplitude at $5\%$ or $10\%$.
The first three contributions are evaluated by local linear propagation,
while the phase contribution uses the second-order small-phase propagation
in Eq.~\eqref{sm:phase-uncertainty} over $0\le\sigma_\Phi\le8^\circ$.

\FloatBarrier

\begin{thebibliography}{48}\makeatletter
\providecommand \@ifxundefined [1]{ \@ifx{#1\undefined}
}\providecommand \@ifnum [1]{ \ifnum #1\expandafter \@firstoftwo
 \else \expandafter \@secondoftwo
 \fi
}\providecommand \@ifx [1]{ \ifx #1\expandafter \@firstoftwo
 \else \expandafter \@secondoftwo
 \fi
}\providecommand \natexlab [1]{#1}\providecommand \enquote  [1]{``#1''}\providecommand \bibnamefont  [1]{#1}\providecommand \bibfnamefont [1]{#1}\providecommand \citenamefont [1]{#1}\providecommand \href@noop [0]{\@secondoftwo}\providecommand \href [0]{\begingroup \@sanitize@url \@href}\providecommand \@href[1]{\@@startlink{#1}\@@href}\providecommand \@@href[1]{\endgroup#1\@@endlink}\providecommand \@sanitize@url [0]{\catcode `\\12\catcode `\$12\catcode
  `\&12\catcode `\#12\catcode `\^12\catcode `\_12\catcode `\%12\relax}\providecommand \@@startlink[1]{}\providecommand \@@endlink[0]{}\providecommand \url  [0]{\begingroup\@sanitize@url \@url }\providecommand \@url [1]{\endgroup\@href {#1}{\urlprefix }}\providecommand \urlprefix  [0]{URL }\providecommand \Eprint [0]{\href }\providecommand \doibase [0]{https://doi.org/}\providecommand \selectlanguage [0]{\@gobble}\providecommand \bibinfo  [0]{\@secondoftwo}\providecommand \bibfield  [0]{\@secondoftwo}\providecommand \translation [1]{[#1]}\providecommand \BibitemOpen [0]{}\providecommand \bibitemStop [0]{}\providecommand \bibitemNoStop [0]{.\EOS\space}\providecommand \EOS [0]{\spacefactor3000\relax}\providecommand \BibitemShut  [1]{\csname bibitem#1\endcsname}\let\auto@bib@innerbib\@empty
\bibitem [{\citenamefont {Tiedau}\ \emph {et~al.}(2024)\citenamefont {Tiedau},
  \citenamefont {Okhapkin}, \citenamefont {Zhang}, \citenamefont {Thielking},
  \citenamefont {Zitzer}, \citenamefont {Peik}, \citenamefont {Schaden},
  \citenamefont {Pronebner}, \citenamefont {Morawetz}, \citenamefont {Toscani
  De~Col}, \citenamefont {Schneider}, \citenamefont {Leitner}, \citenamefont
  {Pressler}, \citenamefont {Kazakov}, \citenamefont {Beeks}, \citenamefont
  {Sikorsky},\ and\ \citenamefont {Schumm}}]{Tiedau2024}  \BibitemOpen
  \bibfield  {author} {\bibinfo {author} {\bibfnamefont {J.}~\bibnamefont
  {Tiedau}}, \bibinfo {author} {\bibfnamefont {M.~V.}\ \bibnamefont
  {Okhapkin}}, \bibinfo {author} {\bibfnamefont {K.}~\bibnamefont {Zhang}},
  \bibinfo {author} {\bibfnamefont {J.}~\bibnamefont {Thielking}}, \bibinfo
  {author} {\bibfnamefont {G.}~\bibnamefont {Zitzer}}, \bibinfo {author}
  {\bibfnamefont {E.}~\bibnamefont {Peik}}, \bibinfo {author} {\bibfnamefont
  {F.}~\bibnamefont {Schaden}}, \bibinfo {author} {\bibfnamefont
  {T.}~\bibnamefont {Pronebner}}, \bibinfo {author} {\bibfnamefont
  {I.}~\bibnamefont {Morawetz}}, \bibinfo {author} {\bibfnamefont
  {L.}~\bibnamefont {Toscani De~Col}}, \bibinfo {author} {\bibfnamefont
  {F.}~\bibnamefont {Schneider}}, \bibinfo {author} {\bibfnamefont
  {A.}~\bibnamefont {Leitner}}, \bibinfo {author} {\bibfnamefont
  {M.}~\bibnamefont {Pressler}}, \bibinfo {author} {\bibfnamefont {G.~A.}\
  \bibnamefont {Kazakov}}, \bibinfo {author} {\bibfnamefont {K.}~\bibnamefont
  {Beeks}}, \bibinfo {author} {\bibfnamefont {T.}~\bibnamefont {Sikorsky}},\
  and\ \bibinfo {author} {\bibfnamefont {T.}~\bibnamefont {Schumm}},\
  }\bibfield  {title} {\bibinfo {title} {Laser excitation of the {Th-229}
  nucleus},\ }\href {https://doi.org/10.1103/PhysRevLett.132.182501} {\bibfield
   {journal} {\bibinfo  {journal} {Phys. Rev. Lett.}\ }\textbf {\bibinfo
  {volume} {132}}(\bibinfo {number} {18}),\ \bibinfo {pages} {182501} (\bibinfo
  {year} {2024})}\BibitemShut {NoStop}\bibitem [{\citenamefont {Elwell}\ \emph {et~al.}(2024)\citenamefont {Elwell},
  \citenamefont {Schneider}, \citenamefont {Jeet}, \citenamefont {Terhune},
  \citenamefont {Morgan}, \citenamefont {Alexandrova}, \citenamefont
  {Tran~Tan}, \citenamefont {Derevianko},\ and\ \citenamefont
  {Hudson}}]{Elwell2024}  \BibitemOpen
  \bibfield  {author} {\bibinfo {author} {\bibfnamefont {R.}~\bibnamefont
  {Elwell}}, \bibinfo {author} {\bibfnamefont {C.}~\bibnamefont {Schneider}},
  \bibinfo {author} {\bibfnamefont {J.}~\bibnamefont {Jeet}}, \bibinfo {author}
  {\bibfnamefont {J.~E.~S.}\ \bibnamefont {Terhune}}, \bibinfo {author}
  {\bibfnamefont {H.~W.~T.}\ \bibnamefont {Morgan}}, \bibinfo {author}
  {\bibfnamefont {A.~N.}\ \bibnamefont {Alexandrova}}, \bibinfo {author}
  {\bibfnamefont {H.~B.}\ \bibnamefont {Tran~Tan}}, \bibinfo {author}
  {\bibfnamefont {A.}~\bibnamefont {Derevianko}},\ and\ \bibinfo {author}
  {\bibfnamefont {E.~R.}\ \bibnamefont {Hudson}},\ }\bibfield  {title}
  {\bibinfo {title} {Laser excitation of the {$^{229}$Th} nuclear isomeric
  transition in a solid-state host},\ }\href
  {https://doi.org/10.1103/PhysRevLett.133.013201} {\bibfield  {journal}
  {\bibinfo  {journal} {Phys. Rev. Lett.}\ }\textbf {\bibinfo {volume}
  {133}}(\bibinfo {number} {1}),\ \bibinfo {pages} {013201} (\bibinfo {year}
  {2024})}\BibitemShut {NoStop}\bibitem [{\citenamefont {Zhang}\ \emph
  {et~al.}(2024{\natexlab{a}})\citenamefont {Zhang}, \citenamefont {von~der
  Wense}, \citenamefont {Doyle}, \citenamefont {Higgins}, \citenamefont {Ooi},
  \citenamefont {Friebel}, \citenamefont {Ye}, \citenamefont {Elwell},
  \citenamefont {Terhune}, \citenamefont {Morgan}, \citenamefont {Alexandrova},
  \citenamefont {Tran~Tan}, \citenamefont {Derevianko},\ and\ \citenamefont
  {Hudson}}]{ZhangFilms2024}  \BibitemOpen
  \bibfield  {author} {\bibinfo {author} {\bibfnamefont {C.}~\bibnamefont
  {Zhang}}, \bibinfo {author} {\bibfnamefont {L.}~\bibnamefont {von~der
  Wense}}, \bibinfo {author} {\bibfnamefont {J.~F.}\ \bibnamefont {Doyle}},
  \bibinfo {author} {\bibfnamefont {J.~S.}\ \bibnamefont {Higgins}}, \bibinfo
  {author} {\bibfnamefont {T.}~\bibnamefont {Ooi}}, \bibinfo {author}
  {\bibfnamefont {H.~U.}\ \bibnamefont {Friebel}}, \bibinfo {author}
  {\bibfnamefont {J.}~\bibnamefont {Ye}}, \bibinfo {author} {\bibfnamefont
  {R.}~\bibnamefont {Elwell}}, \bibinfo {author} {\bibfnamefont {J.~E.~S.}\
  \bibnamefont {Terhune}}, \bibinfo {author} {\bibfnamefont {H.~W.~T.}\
  \bibnamefont {Morgan}}, \bibinfo {author} {\bibfnamefont {A.~N.}\
  \bibnamefont {Alexandrova}}, \bibinfo {author} {\bibfnamefont {H.~B.}\
  \bibnamefont {Tran~Tan}}, \bibinfo {author} {\bibfnamefont {A.}~\bibnamefont
  {Derevianko}},\ and\ \bibinfo {author} {\bibfnamefont {E.~R.}\ \bibnamefont
  {Hudson}},\ }\bibfield  {title} {\bibinfo {title} {{$^{229}$ThF$_4$} thin
  films for solid-state nuclear clocks},\ }\href
  {https://doi.org/10.1038/s41586-024-08256-5} {\bibfield  {journal} {\bibinfo
  {journal} {Nature}\ }\textbf {\bibinfo {volume} {636}}(\bibinfo {number}
  {8043}),\ \bibinfo {pages} {603--608} (\bibinfo {year}
  {2024}{\natexlab{a}})}\BibitemShut {NoStop}\bibitem [{\citenamefont {Elwell}\ \emph {et~al.}(2025)\citenamefont {Elwell},
  \citenamefont {Terhune}, \citenamefont {Schneider}, \citenamefont {Morgan},
  \citenamefont {Tran~Tan}, \citenamefont {Perera}, \citenamefont {Rehn},
  \citenamefont {Alfonso}, \citenamefont {von~der Wense}, \citenamefont
  {Seiferle}, \citenamefont {Scharl}, \citenamefont {Thirolf}, \citenamefont
  {Derevianko},\ and\ \citenamefont {Hudson}}]{Elwell2025}  \BibitemOpen
  \bibfield  {author} {\bibinfo {author} {\bibfnamefont {R.}~\bibnamefont
  {Elwell}}, \bibinfo {author} {\bibfnamefont {J.~E.~S.}\ \bibnamefont
  {Terhune}}, \bibinfo {author} {\bibfnamefont {C.}~\bibnamefont {Schneider}},
  \bibinfo {author} {\bibfnamefont {H.~W.~T.}\ \bibnamefont {Morgan}}, \bibinfo
  {author} {\bibfnamefont {H.~B.}\ \bibnamefont {Tran~Tan}}, \bibinfo {author}
  {\bibfnamefont {U.~C.}\ \bibnamefont {Perera}}, \bibinfo {author}
  {\bibfnamefont {D.~A.}\ \bibnamefont {Rehn}}, \bibinfo {author}
  {\bibfnamefont {M.~C.}\ \bibnamefont {Alfonso}}, \bibinfo {author}
  {\bibfnamefont {L.}~\bibnamefont {von~der Wense}}, \bibinfo {author}
  {\bibfnamefont {B.}~\bibnamefont {Seiferle}}, \bibinfo {author}
  {\bibfnamefont {K.}~\bibnamefont {Scharl}}, \bibinfo {author} {\bibfnamefont
  {P.~G.}\ \bibnamefont {Thirolf}}, \bibinfo {author} {\bibfnamefont
  {A.}~\bibnamefont {Derevianko}},\ and\ \bibinfo {author} {\bibfnamefont
  {E.~R.}\ \bibnamefont {Hudson}},\ }\bibfield  {title} {\bibinfo {title}
  {Laser-based conversion electron {M\"ossbauer} spectroscopy of
  {$^{229}$ThO$_2$}},\ }\href {https://doi.org/10.1038/s41586-025-09776-4}
  {\bibfield  {journal} {\bibinfo  {journal} {Nature}\ }\textbf {\bibinfo
  {volume} {648}}(\bibinfo {number} {8093}),\ \bibinfo {pages} {300--305}
  (\bibinfo {year} {2025})}\BibitemShut {NoStop}\bibitem [{\citenamefont {Morawetz}\ \emph {et~al.}(2026)\citenamefont
  {Morawetz}, \citenamefont {Riebner}, \citenamefont {Toscani De~Col},
  \citenamefont {Schneider}, \citenamefont {Sempelmann}, \citenamefont
  {Schaden}, \citenamefont {Bartokos}, \citenamefont {Kazakov}, \citenamefont
  {Lahs}, \citenamefont {Beeks}, \citenamefont {Gerstenecker}, \citenamefont
  {Gr{\"u}neis}, \citenamefont {Pimon}, \citenamefont {Schumm}, \citenamefont
  {Lal}, \citenamefont {Zitzer}, \citenamefont {Petrov}, \citenamefont
  {Tiedau}, \citenamefont {Okhapkin},\ and\ \citenamefont
  {Peik}}]{Morawetz2026}  \BibitemOpen
  \bibfield  {author} {\bibinfo {author} {\bibfnamefont {I.}~\bibnamefont
  {Morawetz}}, \bibinfo {author} {\bibfnamefont {T.}~\bibnamefont {Riebner}},
  \bibinfo {author} {\bibfnamefont {L.}~\bibnamefont {Toscani De~Col}},
  \bibinfo {author} {\bibfnamefont {F.}~\bibnamefont {Schneider}}, \bibinfo
  {author} {\bibfnamefont {N.}~\bibnamefont {Sempelmann}}, \bibinfo {author}
  {\bibfnamefont {F.}~\bibnamefont {Schaden}}, \bibinfo {author} {\bibfnamefont
  {M.}~\bibnamefont {Bartokos}}, \bibinfo {author} {\bibfnamefont {G.~A.}\
  \bibnamefont {Kazakov}}, \bibinfo {author} {\bibfnamefont {S.}~\bibnamefont
  {Lahs}}, \bibinfo {author} {\bibfnamefont {K.}~\bibnamefont {Beeks}},
  \bibinfo {author} {\bibfnamefont {B.}~\bibnamefont {Gerstenecker}}, \bibinfo
  {author} {\bibfnamefont {A.}~\bibnamefont {Gr{\"u}neis}}, \bibinfo {author}
  {\bibfnamefont {M.}~\bibnamefont {Pimon}}, \bibinfo {author} {\bibfnamefont
  {T.}~\bibnamefont {Schumm}}, \bibinfo {author} {\bibfnamefont
  {V.}~\bibnamefont {Lal}}, \bibinfo {author} {\bibfnamefont {G.}~\bibnamefont
  {Zitzer}}, \bibinfo {author} {\bibfnamefont {V.}~\bibnamefont {Petrov}},
  \bibinfo {author} {\bibfnamefont {J.}~\bibnamefont {Tiedau}}, \bibinfo
  {author} {\bibfnamefont {M.~V.}\ \bibnamefont {Okhapkin}},\ and\ \bibinfo
  {author} {\bibfnamefont {E.}~\bibnamefont {Peik}},\ }\bibfield  {title}
  {\bibinfo {title} {Continuous-wave laser absorption spectroscopy of the
  thorium-229 nucleus},\ }\href {https://doi.org/10.1038/s41586-026-11011-7}
  {\bibfield  {journal} {\bibinfo  {journal} {Nature}\ }\textbf {\bibinfo
  {volume} {657}}(\bibinfo {number} {8132}),\ \bibinfo {pages} {626--631}
  (\bibinfo {year} {2026})}\BibitemShut {NoStop}\bibitem [{\citenamefont {Zhang}\ \emph
  {et~al.}(2024{\natexlab{b}})\citenamefont {Zhang}, \citenamefont {Ooi},
  \citenamefont {Higgins}, \citenamefont {Doyle}, \citenamefont {von~der
  Wense}, \citenamefont {Beeks}, \citenamefont {Leitner}, \citenamefont
  {Kazakov}, \citenamefont {Li}, \citenamefont {Thirolf}, \citenamefont
  {Schumm},\ and\ \citenamefont {Ye}}]{Zhang2024}  \BibitemOpen
  \bibfield  {author} {\bibinfo {author} {\bibfnamefont {C.}~\bibnamefont
  {Zhang}}, \bibinfo {author} {\bibfnamefont {T.}~\bibnamefont {Ooi}}, \bibinfo
  {author} {\bibfnamefont {J.~S.}\ \bibnamefont {Higgins}}, \bibinfo {author}
  {\bibfnamefont {J.~F.}\ \bibnamefont {Doyle}}, \bibinfo {author}
  {\bibfnamefont {L.}~\bibnamefont {von~der Wense}}, \bibinfo {author}
  {\bibfnamefont {K.}~\bibnamefont {Beeks}}, \bibinfo {author} {\bibfnamefont
  {A.}~\bibnamefont {Leitner}}, \bibinfo {author} {\bibfnamefont {G.~A.}\
  \bibnamefont {Kazakov}}, \bibinfo {author} {\bibfnamefont {P.}~\bibnamefont
  {Li}}, \bibinfo {author} {\bibfnamefont {P.~G.}\ \bibnamefont {Thirolf}},
  \bibinfo {author} {\bibfnamefont {T.}~\bibnamefont {Schumm}},\ and\ \bibinfo
  {author} {\bibfnamefont {J.}~\bibnamefont {Ye}},\ }\bibfield  {title}
  {\bibinfo {title} {Frequency ratio of the {$^{229\mathrm{m}}$Th} nuclear
  isomeric transition and the {$^{87}$Sr} atomic clock},\ }\href
  {https://doi.org/10.1038/s41586-024-07839-6} {\bibfield  {journal} {\bibinfo
  {journal} {Nature}\ }\textbf {\bibinfo {volume} {633}}(\bibinfo {number}
  {8028}),\ \bibinfo {pages} {63--70} (\bibinfo {year}
  {2024}{\natexlab{b}})}\BibitemShut {NoStop}\bibitem [{\citenamefont {Higgins}\ \emph {et~al.}(2025)\citenamefont
  {Higgins}, \citenamefont {Ooi}, \citenamefont {Doyle}, \citenamefont {Zhang},
  \citenamefont {Ye}, \citenamefont {Beeks}, \citenamefont {Sikorsky},\ and\
  \citenamefont {Schumm}}]{Higgins2025}  \BibitemOpen
  \bibfield  {author} {\bibinfo {author} {\bibfnamefont {J.~S.}\ \bibnamefont
  {Higgins}}, \bibinfo {author} {\bibfnamefont {T.}~\bibnamefont {Ooi}},
  \bibinfo {author} {\bibfnamefont {J.~F.}\ \bibnamefont {Doyle}}, \bibinfo
  {author} {\bibfnamefont {C.}~\bibnamefont {Zhang}}, \bibinfo {author}
  {\bibfnamefont {J.}~\bibnamefont {Ye}}, \bibinfo {author} {\bibfnamefont
  {K.}~\bibnamefont {Beeks}}, \bibinfo {author} {\bibfnamefont
  {T.}~\bibnamefont {Sikorsky}},\ and\ \bibinfo {author} {\bibfnamefont
  {T.}~\bibnamefont {Schumm}},\ }\bibfield  {title} {\bibinfo {title}
  {Temperature sensitivity of a thorium-229 solid-state nuclear clock},\ }\href
  {https://doi.org/10.1103/PhysRevLett.134.113801} {\bibfield  {journal}
  {\bibinfo  {journal} {Phys. Rev. Lett.}\ }\textbf {\bibinfo {volume}
  {134}}(\bibinfo {number} {11}),\ \bibinfo {pages} {113801} (\bibinfo {year}
  {2025})}\BibitemShut {NoStop}\bibitem [{\citenamefont {Ooi}\ \emph {et~al.}(2026)\citenamefont {Ooi},
  \citenamefont {Doyle}, \citenamefont {Zhang}, \citenamefont {Higgins},
  \citenamefont {Ye}, \citenamefont {Beeks}, \citenamefont {Sikorsky},\ and\
  \citenamefont {Schumm}}]{Ooi2026}  \BibitemOpen
  \bibfield  {author} {\bibinfo {author} {\bibfnamefont {T.}~\bibnamefont
  {Ooi}}, \bibinfo {author} {\bibfnamefont {J.~F.}\ \bibnamefont {Doyle}},
  \bibinfo {author} {\bibfnamefont {C.}~\bibnamefont {Zhang}}, \bibinfo
  {author} {\bibfnamefont {J.~S.}\ \bibnamefont {Higgins}}, \bibinfo {author}
  {\bibfnamefont {J.}~\bibnamefont {Ye}}, \bibinfo {author} {\bibfnamefont
  {K.}~\bibnamefont {Beeks}}, \bibinfo {author} {\bibfnamefont
  {T.}~\bibnamefont {Sikorsky}},\ and\ \bibinfo {author} {\bibfnamefont
  {T.}~\bibnamefont {Schumm}},\ }\bibfield  {title} {\bibinfo {title}
  {Frequency reproducibility of solid-state thorium-229 nuclear clocks},\
  }\href {https://doi.org/10.1038/s41586-025-09999-5} {\bibfield  {journal}
  {\bibinfo  {journal} {Nature}\ }\textbf {\bibinfo {volume} {650}}(\bibinfo
  {number} {8100}),\ \bibinfo {pages} {72--78} (\bibinfo {year}
  {2026})}\BibitemShut {NoStop}\bibitem [{\citenamefont {Toscani De~Col}\ \emph {et~al.}(2026)\citenamefont
  {Toscani De~Col}, \citenamefont {Riebner}, \citenamefont {Morawetz},
  \citenamefont {Schneider}, \citenamefont {Sempelmann}, \citenamefont
  {Schlachet-L{\'e}pinay}, \citenamefont {Schaden}, \citenamefont {Bartokos},
  \citenamefont {Kazakov}, \citenamefont {Beeks}, \citenamefont {Gerstenecker},
  \citenamefont {Pimon}, \citenamefont {Lahs}, \citenamefont {Hellerschmied},
  \citenamefont {Lercher}, \citenamefont {Denker}, \citenamefont {Premper},
  \citenamefont {Niessner}, \citenamefont {Matus}, \citenamefont {{\v
  C}{\'i}žek}, \citenamefont {{\v C}{\'i}p}, \citenamefont {Lal},
  \citenamefont {Zitzer}, \citenamefont {Petrov}, \citenamefont {Tiedau},
  \citenamefont {Okhapkin}, \citenamefont {Peik},\ and\ \citenamefont
  {Schumm}}]{ToscaniClock2026}  \BibitemOpen
  \bibfield  {author} {\bibinfo {author} {\bibfnamefont {L.}~\bibnamefont
  {Toscani De~Col}}, \bibinfo {author} {\bibfnamefont {T.}~\bibnamefont
  {Riebner}}, \bibinfo {author} {\bibfnamefont {I.}~\bibnamefont {Morawetz}},
  \bibinfo {author} {\bibfnamefont {F.}~\bibnamefont {Schneider}}, \bibinfo
  {author} {\bibfnamefont {N.}~\bibnamefont {Sempelmann}}, \bibinfo {author}
  {\bibfnamefont {J.}~\bibnamefont {Schlachet-L{\'e}pinay}}, \bibinfo {author}
  {\bibfnamefont {F.}~\bibnamefont {Schaden}}, \bibinfo {author} {\bibfnamefont
  {M.}~\bibnamefont {Bartokos}}, \bibinfo {author} {\bibfnamefont {G.~A.}\
  \bibnamefont {Kazakov}}, \bibinfo {author} {\bibfnamefont {K.}~\bibnamefont
  {Beeks}}, \bibinfo {author} {\bibfnamefont {B.}~\bibnamefont {Gerstenecker}},
  \bibinfo {author} {\bibfnamefont {M.}~\bibnamefont {Pimon}}, \bibinfo
  {author} {\bibfnamefont {S.}~\bibnamefont {Lahs}}, \bibinfo {author}
  {\bibfnamefont {A.}~\bibnamefont {Hellerschmied}}, \bibinfo {author}
  {\bibfnamefont {T.}~\bibnamefont {Lercher}}, \bibinfo {author} {\bibfnamefont
  {H.}~\bibnamefont {Denker}}, \bibinfo {author} {\bibfnamefont
  {J.}~\bibnamefont {Premper}}, \bibinfo {author} {\bibfnamefont
  {A.}~\bibnamefont {Niessner}}, \bibinfo {author} {\bibfnamefont
  {M.}~\bibnamefont {Matus}}, \bibinfo {author} {\bibfnamefont
  {M.}~\bibnamefont {{\v C}{\'i}žek}}, \bibinfo {author} {\bibfnamefont
  {O.}~\bibnamefont {{\v C}{\'i}p}}, \bibinfo {author} {\bibfnamefont
  {V.}~\bibnamefont {Lal}}, \bibinfo {author} {\bibfnamefont {G.}~\bibnamefont
  {Zitzer}}, \bibinfo {author} {\bibfnamefont {V.}~\bibnamefont {Petrov}},
  \bibinfo {author} {\bibfnamefont {J.}~\bibnamefont {Tiedau}}, \bibinfo
  {author} {\bibfnamefont {M.~V.}\ \bibnamefont {Okhapkin}}, \bibinfo {author}
  {\bibfnamefont {E.}~\bibnamefont {Peik}},\ and\ \bibinfo {author}
  {\bibfnamefont {T.}~\bibnamefont {Schumm}},\ }\bibfield  {title} {\bibinfo
  {title} {A thorium-229 optical nuclear clock with feedback loop},\ }\href
  {https://doi.org/10.1038/s41586-026-11084-4} {\bibfield  {journal} {\bibinfo
  {journal} {Nature}\ }\textbf {\bibinfo {volume} {658}},\ \bibinfo {pages}
  {897--902} (\bibinfo {year} {2026})}\BibitemShut {NoStop}\bibitem [{\citenamefont {Huang}\ \emph {et~al.}(2026)\citenamefont {Huang},
  \citenamefont {Yan}, \citenamefont {Xiao}, \citenamefont {Bu}, \citenamefont
  {Zhao}, \citenamefont {Zhang}, \citenamefont {Yan}, \citenamefont {Chen},
  \citenamefont {Zhang}, \citenamefont {Penyazkov}, \citenamefont {Zhan},
  \citenamefont {Yan}, \citenamefont {Wang}, \citenamefont {Li}, \citenamefont
  {Li}, \citenamefont {Jiang}, \citenamefont {Qian}, \citenamefont {Liu},
  \citenamefont {He}, \citenamefont {Sun}, \citenamefont {Tian}, \citenamefont
  {Lu}, \citenamefont {Ma}, \citenamefont {Li}, \citenamefont {Wu},
  \citenamefont {Gong}, \citenamefont {Li}, \citenamefont {Shi}, \citenamefont
  {Li}, \citenamefont {Ma}, \citenamefont {Zhu}, \citenamefont {Mo},
  \citenamefont {Lin}, \citenamefont {You}, \citenamefont {Lin}, \citenamefont
  {Zhang}, \citenamefont {Hang}, \citenamefont {Su},\ and\ \citenamefont
  {Ding}}]{HuangClock2026}  \BibitemOpen
  \bibfield  {author} {\bibinfo {author} {\bibfnamefont {B.}~\bibnamefont
  {Huang}}, \bibinfo {author} {\bibfnamefont {G.}~\bibnamefont {Yan}}, \bibinfo
  {author} {\bibfnamefont {Q.}~\bibnamefont {Xiao}}, \bibinfo {author}
  {\bibfnamefont {W.}~\bibnamefont {Bu}}, \bibinfo {author} {\bibfnamefont
  {C.}~\bibnamefont {Zhao}}, \bibinfo {author} {\bibfnamefont {Z.}~\bibnamefont
  {Zhang}}, \bibinfo {author} {\bibfnamefont {C.}~\bibnamefont {Yan}}, \bibinfo
  {author} {\bibfnamefont {Z.-A.}\ \bibnamefont {Chen}}, \bibinfo {author}
  {\bibfnamefont {P.}~\bibnamefont {Zhang}}, \bibinfo {author} {\bibfnamefont
  {G.}~\bibnamefont {Penyazkov}}, \bibinfo {author} {\bibfnamefont
  {Z.}~\bibnamefont {Zhan}}, \bibinfo {author} {\bibfnamefont {L.}~\bibnamefont
  {Yan}}, \bibinfo {author} {\bibfnamefont {Y.}~\bibnamefont {Wang}}, \bibinfo
  {author} {\bibfnamefont {L.}~\bibnamefont {Li}}, \bibinfo {author}
  {\bibfnamefont {S.}~\bibnamefont {Li}}, \bibinfo {author} {\bibfnamefont
  {D.}~\bibnamefont {Jiang}}, \bibinfo {author} {\bibfnamefont
  {X.}~\bibnamefont {Qian}}, \bibinfo {author} {\bibfnamefont {X.}~\bibnamefont
  {Liu}}, \bibinfo {author} {\bibfnamefont {Q.}~\bibnamefont {He}}, \bibinfo
  {author} {\bibfnamefont {T.}~\bibnamefont {Sun}}, \bibinfo {author}
  {\bibfnamefont {H.}~\bibnamefont {Tian}}, \bibinfo {author} {\bibfnamefont
  {B.}~\bibnamefont {Lu}}, \bibinfo {author} {\bibfnamefont {N.}~\bibnamefont
  {Ma}}, \bibinfo {author} {\bibfnamefont {J.}~\bibnamefont {Li}}, \bibinfo
  {author} {\bibfnamefont {Y.}~\bibnamefont {Wu}}, \bibinfo {author}
  {\bibfnamefont {Q.}~\bibnamefont {Gong}}, \bibinfo {author} {\bibfnamefont
  {Y.}~\bibnamefont {Li}}, \bibinfo {author} {\bibfnamefont {H.}~\bibnamefont
  {Shi}}, \bibinfo {author} {\bibfnamefont {X.}~\bibnamefont {Li}}, \bibinfo
  {author} {\bibfnamefont {L.}~\bibnamefont {Ma}}, \bibinfo {author}
  {\bibfnamefont {S.}~\bibnamefont {Zhu}}, \bibinfo {author} {\bibfnamefont
  {Y.}~\bibnamefont {Mo}}, \bibinfo {author} {\bibfnamefont {J.}~\bibnamefont
  {Lin}}, \bibinfo {author} {\bibfnamefont {L.}~\bibnamefont {You}}, \bibinfo
  {author} {\bibfnamefont {Y.}~\bibnamefont {Lin}}, \bibinfo {author}
  {\bibfnamefont {X.}~\bibnamefont {Zhang}}, \bibinfo {author} {\bibfnamefont
  {Y.}~\bibnamefont {Hang}}, \bibinfo {author} {\bibfnamefont {L.}~\bibnamefont
  {Su}},\ and\ \bibinfo {author} {\bibfnamefont {S.}~\bibnamefont {Ding}},\
  }\bibfield  {title} {\bibinfo {title} {A nuclear clock synchronized to
  {$^{229}$Th}},\ }\href {https://doi.org/10.1038/s41586-026-11122-1}
  {\bibfield  {journal} {\bibinfo  {journal} {Nature}\ }\textbf {\bibinfo
  {volume} {658}},\ \bibinfo {pages} {903--908} (\bibinfo {year}
  {2026})}\BibitemShut {NoStop}\bibitem [{\citenamefont {Flambaum}(2006)}]{Flambaum2006}  \BibitemOpen
  \bibfield  {author} {\bibinfo {author} {\bibfnamefont {V.~V.}\ \bibnamefont
  {Flambaum}},\ }\bibfield  {title} {\bibinfo {title} {Enhanced effect of
  temporal variation of the fine structure constant and the strong interaction
  in {$^{229}$Th}},\ }\href {https://doi.org/10.1103/physrevlett.97.092502}
  {\bibfield  {journal} {\bibinfo  {journal} {Phys. Rev. Lett.}\ }\textbf
  {\bibinfo {volume} {97}}(\bibinfo {number} {9}),\ \bibinfo {pages} {092502}
  (\bibinfo {year} {2006})}\BibitemShut {NoStop}\bibitem [{\citenamefont {Fadeev}\ \emph {et~al.}(2020)\citenamefont {Fadeev},
  \citenamefont {Berengut},\ and\ \citenamefont {Flambaum}}]{Fadeev2020}  \BibitemOpen
  \bibfield  {author} {\bibinfo {author} {\bibfnamefont {P.}~\bibnamefont
  {Fadeev}}, \bibinfo {author} {\bibfnamefont {J.~C.}\ \bibnamefont
  {Berengut}},\ and\ \bibinfo {author} {\bibfnamefont {V.~V.}\ \bibnamefont
  {Flambaum}},\ }\bibfield  {title} {\bibinfo {title} {Sensitivity of
  {$^{229}$Th} nuclear clock transition to variation of the fine-structure
  constant},\ }\href {https://doi.org/10.1103/physreva.102.052833} {\bibfield
  {journal} {\bibinfo  {journal} {Phys. Rev. A}\ }\textbf {\bibinfo {volume}
  {102}}(\bibinfo {number} {5}),\ \bibinfo {pages} {052833} (\bibinfo {year}
  {2020})}\BibitemShut {NoStop}\bibitem [{\citenamefont {Yamaguchi}\ \emph {et~al.}(2024)\citenamefont
  {Yamaguchi}, \citenamefont {Shigekawa}, \citenamefont {Haba}, \citenamefont
  {Kikunaga}, \citenamefont {Shirasaki}, \citenamefont {Wada},\ and\
  \citenamefont {Katori}}]{Yamaguchi2024}  \BibitemOpen
  \bibfield  {author} {\bibinfo {author} {\bibfnamefont {A.}~\bibnamefont
  {Yamaguchi}}, \bibinfo {author} {\bibfnamefont {Y.}~\bibnamefont
  {Shigekawa}}, \bibinfo {author} {\bibfnamefont {H.}~\bibnamefont {Haba}},
  \bibinfo {author} {\bibfnamefont {H.}~\bibnamefont {Kikunaga}}, \bibinfo
  {author} {\bibfnamefont {K.}~\bibnamefont {Shirasaki}}, \bibinfo {author}
  {\bibfnamefont {M.}~\bibnamefont {Wada}},\ and\ \bibinfo {author}
  {\bibfnamefont {H.}~\bibnamefont {Katori}},\ }\bibfield  {title} {\bibinfo
  {title} {Laser spectroscopy of triply charged {$^{229}$Th} isomer for a
  nuclear clock},\ }\href {https://doi.org/10.1038/s41586-024-07296-1}
  {\bibfield  {journal} {\bibinfo  {journal} {Nature}\ }\textbf {\bibinfo
  {volume} {629}}(\bibinfo {number} {8010}),\ \bibinfo {pages} {62--66}
  (\bibinfo {year} {2024})}\BibitemShut {NoStop}\bibitem [{\citenamefont {Beeks}\ \emph {et~al.}(2025)\citenamefont {Beeks},
  \citenamefont {Kazakov}, \citenamefont {Schaden}, \citenamefont {Morawetz},
  \citenamefont {Toscani De~Col}, \citenamefont {Riebner}, \citenamefont
  {Bartokos}, \citenamefont {Sikorsky}, \citenamefont {Schumm}, \citenamefont
  {Zhang}, \citenamefont {Ooi}, \citenamefont {Higgins}, \citenamefont {Doyle},
  \citenamefont {Ye},\ and\ \citenamefont {Safronova}}]{Beeks2025}  \BibitemOpen
  \bibfield  {author} {\bibinfo {author} {\bibfnamefont {K.}~\bibnamefont
  {Beeks}}, \bibinfo {author} {\bibfnamefont {G.~A.}\ \bibnamefont {Kazakov}},
  \bibinfo {author} {\bibfnamefont {F.}~\bibnamefont {Schaden}}, \bibinfo
  {author} {\bibfnamefont {I.}~\bibnamefont {Morawetz}}, \bibinfo {author}
  {\bibfnamefont {L.}~\bibnamefont {Toscani De~Col}}, \bibinfo {author}
  {\bibfnamefont {T.}~\bibnamefont {Riebner}}, \bibinfo {author} {\bibfnamefont
  {M.}~\bibnamefont {Bartokos}}, \bibinfo {author} {\bibfnamefont
  {T.}~\bibnamefont {Sikorsky}}, \bibinfo {author} {\bibfnamefont
  {T.}~\bibnamefont {Schumm}}, \bibinfo {author} {\bibfnamefont
  {C.}~\bibnamefont {Zhang}}, \bibinfo {author} {\bibfnamefont
  {T.}~\bibnamefont {Ooi}}, \bibinfo {author} {\bibfnamefont {J.~S.}\
  \bibnamefont {Higgins}}, \bibinfo {author} {\bibfnamefont {J.~F.}\
  \bibnamefont {Doyle}}, \bibinfo {author} {\bibfnamefont {J.}~\bibnamefont
  {Ye}},\ and\ \bibinfo {author} {\bibfnamefont {M.~S.}\ \bibnamefont
  {Safronova}},\ }\bibfield  {title} {\bibinfo {title} {Fine-structure constant
  sensitivity of the th-229 nuclear clock transition},\ }\href
  {https://doi.org/10.1038/s41467-025-64191-7} {\bibfield  {journal} {\bibinfo
  {journal} {Nat. Commun.}\ }\textbf {\bibinfo {volume} {16}}(\bibinfo {number}
  {1}),\ \bibinfo {pages} {9147} (\bibinfo {year} {2025})}\BibitemShut
  {NoStop}\bibitem [{\citenamefont {Ruchowska}\ \emph {et~al.}(2006)\citenamefont
  {Ruchowska}, \citenamefont {P{\l}{\'o}ciennik}, \citenamefont {{\.Z}ylicz},
  \citenamefont {Mach}, \citenamefont {Kvasil}, \citenamefont {Algora},
  \citenamefont {Amzal}, \citenamefont {B{\"a}ck}, \citenamefont {Borge},
  \citenamefont {Boutami}, \citenamefont {Butler}, \citenamefont
  {Cederk{\"a}ll}, \citenamefont {Cederwall}, \citenamefont {Fogelberg},
  \citenamefont {Fraile}, \citenamefont {Fynbo}, \citenamefont {Hageb{\o}},
  \citenamefont {Hoff}, \citenamefont {Gausemel}, \citenamefont {Jungclaus},
  \citenamefont {Kaczarowski}, \citenamefont {Kerek}, \citenamefont
  {Kurcewicz}, \citenamefont {Lagergren}, \citenamefont {Nacher}, \citenamefont
  {Rubio}, \citenamefont {Syntfeld}, \citenamefont {Tengblad}, \citenamefont
  {Wasilewski},\ and\ \citenamefont {Weissman}}]{Ruchowska2006}  \BibitemOpen
  \bibfield  {author} {\bibinfo {author} {\bibfnamefont {E.}~\bibnamefont
  {Ruchowska}}, \bibinfo {author} {\bibfnamefont {W.~A.}\ \bibnamefont
  {P{\l}{\'o}ciennik}}, \bibinfo {author} {\bibfnamefont {J.}~\bibnamefont
  {{\.Z}ylicz}}, \bibinfo {author} {\bibfnamefont {H.}~\bibnamefont {Mach}},
  \bibinfo {author} {\bibfnamefont {J.}~\bibnamefont {Kvasil}}, \bibinfo
  {author} {\bibfnamefont {A.}~\bibnamefont {Algora}}, \bibinfo {author}
  {\bibfnamefont {N.}~\bibnamefont {Amzal}}, \bibinfo {author} {\bibfnamefont
  {T.}~\bibnamefont {B{\"a}ck}}, \bibinfo {author} {\bibfnamefont {M.~G.}\
  \bibnamefont {Borge}}, \bibinfo {author} {\bibfnamefont {R.}~\bibnamefont
  {Boutami}}, \bibinfo {author} {\bibfnamefont {P.~A.}\ \bibnamefont {Butler}},
  \bibinfo {author} {\bibfnamefont {J.}~\bibnamefont {Cederk{\"a}ll}}, \bibinfo
  {author} {\bibfnamefont {B.}~\bibnamefont {Cederwall}}, \bibinfo {author}
  {\bibfnamefont {B.}~\bibnamefont {Fogelberg}}, \bibinfo {author}
  {\bibfnamefont {L.~M.}\ \bibnamefont {Fraile}}, \bibinfo {author}
  {\bibfnamefont {H.~O.~U.}\ \bibnamefont {Fynbo}}, \bibinfo {author}
  {\bibfnamefont {E.}~\bibnamefont {Hageb{\o}}}, \bibinfo {author}
  {\bibfnamefont {P.}~\bibnamefont {Hoff}}, \bibinfo {author} {\bibfnamefont
  {H.}~\bibnamefont {Gausemel}}, \bibinfo {author} {\bibfnamefont
  {A.}~\bibnamefont {Jungclaus}}, \bibinfo {author} {\bibfnamefont
  {R.}~\bibnamefont {Kaczarowski}}, \bibinfo {author} {\bibfnamefont
  {A.}~\bibnamefont {Kerek}}, \bibinfo {author} {\bibfnamefont
  {W.}~\bibnamefont {Kurcewicz}}, \bibinfo {author} {\bibfnamefont
  {K.}~\bibnamefont {Lagergren}}, \bibinfo {author} {\bibfnamefont
  {E.}~\bibnamefont {Nacher}}, \bibinfo {author} {\bibfnamefont
  {B.}~\bibnamefont {Rubio}}, \bibinfo {author} {\bibfnamefont
  {A.}~\bibnamefont {Syntfeld}}, \bibinfo {author} {\bibfnamefont
  {O.}~\bibnamefont {Tengblad}}, \bibinfo {author} {\bibfnamefont {A.~A.}\
  \bibnamefont {Wasilewski}},\ and\ \bibinfo {author} {\bibfnamefont
  {L.}~\bibnamefont {Weissman}},\ }\bibfield  {title} {\bibinfo {title}
  {Nuclear structure of {$^{229}$Th}},\ }\href
  {https://doi.org/10.1103/physrevc.73.044326} {\bibfield  {journal} {\bibinfo
  {journal} {Phys. Rev. C}\ }\textbf {\bibinfo {volume} {73}}(\bibinfo {number}
  {4}),\ \bibinfo {pages} {044326} (\bibinfo {year} {2006})}\BibitemShut
  {NoStop}\bibitem [{\citenamefont {Minkov}\ and\ \citenamefont
  {P{\'a}lffy}(2021)}]{Minkov2021}  \BibitemOpen
  \bibfield  {author} {\bibinfo {author} {\bibfnamefont {N.}~\bibnamefont
  {Minkov}}\ and\ \bibinfo {author} {\bibfnamefont {A.}~\bibnamefont
  {P{\'a}lffy}},\ }\bibfield  {title} {\bibinfo {title} {{$^{229\mathrm{m}}$Th}
  isomer from a nuclear model perspective},\ }\href
  {https://doi.org/10.1103/PhysRevC.103.014313} {\bibfield  {journal} {\bibinfo
   {journal} {Phys. Rev. C}\ }\textbf {\bibinfo {volume} {103}}(\bibinfo
  {number} {1}),\ \bibinfo {pages} {014313} (\bibinfo {year}
  {2021})}\BibitemShut {NoStop}\bibitem [{\citenamefont {Minkov}\ and\ \citenamefont
  {P{\'a}lffy}(2017)}]{Minkov2017}  \BibitemOpen
  \bibfield  {author} {\bibinfo {author} {\bibfnamefont {N.}~\bibnamefont
  {Minkov}}\ and\ \bibinfo {author} {\bibfnamefont {A.}~\bibnamefont
  {P{\'a}lffy}},\ }\bibfield  {title} {\bibinfo {title} {Reduced transition
  probabilities for the gamma decay of the 7.8 ev isomer in {$^{229}$Th}},\
  }\href {https://doi.org/10.1103/PhysRevLett.118.212501} {\bibfield  {journal}
  {\bibinfo  {journal} {Phys. Rev. Lett.}\ }\textbf {\bibinfo {volume}
  {118}}(\bibinfo {number} {21}),\ \bibinfo {pages} {212501} (\bibinfo {year}
  {2017})}\BibitemShut {NoStop}\bibitem [{\citenamefont {Zhang}\ \emph
  {et~al.}(2024{\natexlab{c}})\citenamefont {Zhang}, \citenamefont {Li},\ and\
  \citenamefont {Wang}}]{Zhang2024Nonlinear}  \BibitemOpen
  \bibfield  {author} {\bibinfo {author} {\bibfnamefont {H.}~\bibnamefont
  {Zhang}}, \bibinfo {author} {\bibfnamefont {T.}~\bibnamefont {Li}},\ and\
  \bibinfo {author} {\bibfnamefont {X.}~\bibnamefont {Wang}},\ }\bibfield
  {title} {\bibinfo {title} {Highly nonlinear light-nucleus interaction},\
  }\href {https://doi.org/10.1103/PhysRevLett.133.152503} {\bibfield  {journal}
  {\bibinfo  {journal} {Phys. Rev. Lett.}\ }\textbf {\bibinfo {volume} {133}},\
  \bibinfo {pages} {152503} (\bibinfo {year} {2024}{\natexlab{c}})}\BibitemShut
  {NoStop}\bibitem [{\citenamefont {Zhou}\ and\ \citenamefont
  {Wang}(2026)}]{Zhou2026NHM}  \BibitemOpen
  \bibfield  {author} {\bibinfo {author} {\bibfnamefont {J.}~\bibnamefont
  {Zhou}}\ and\ \bibinfo {author} {\bibfnamefont {X.}~\bibnamefont {Wang}},\
  }\bibfield  {title} {\bibinfo {title} {Effects of nuclear hyperfine mixing on
  {$^{229}$Th$^{3+}$} ions},\ }\href {https://doi.org/10.1103/8jhh-ktfy}
  {\bibfield  {journal} {\bibinfo  {journal} {Phys. Rev. Lett.}\ }\textbf
  {\bibinfo {volume} {136}},\ \bibinfo {pages} {092503} (\bibinfo {year}
  {2026})}\BibitemShut {NoStop}\bibitem [{\citenamefont {Bilous}\ \emph {et~al.}(2018)\citenamefont {Bilous},
  \citenamefont {Minkov},\ and\ \citenamefont {P{\'a}lffy}}]{Bilous2018}  \BibitemOpen
  \bibfield  {author} {\bibinfo {author} {\bibfnamefont {P.~V.}\ \bibnamefont
  {Bilous}}, \bibinfo {author} {\bibfnamefont {N.}~\bibnamefont {Minkov}},\
  and\ \bibinfo {author} {\bibfnamefont {A.}~\bibnamefont {P{\'a}lffy}},\
  }\bibfield  {title} {\bibinfo {title} {Electric quadrupole channel of the 7.8
  ev {$^{229}$Th} transition},\ }\href
  {https://doi.org/10.1103/PhysRevC.97.044320} {\bibfield  {journal} {\bibinfo
  {journal} {Phys. Rev. C}\ }\textbf {\bibinfo {volume} {97}}(\bibinfo {number}
  {4}),\ \bibinfo {pages} {044320} (\bibinfo {year} {2018})}\BibitemShut
  {NoStop}\bibitem [{\citenamefont {Kraemer}\ \emph {et~al.}(2023)\citenamefont
  {Kraemer}, \citenamefont {Moens}, \citenamefont {Athanasakis-Kaklamanakis},
  \citenamefont {Bara}, \citenamefont {Beeks}, \citenamefont {Chhetri},
  \citenamefont {Chrysalidis}, \citenamefont {Claessens}, \citenamefont
  {Cocolios}, \citenamefont {Correia}, \citenamefont {De~Witte}, \citenamefont
  {Ferrer}, \citenamefont {Geldhof}, \citenamefont {Heinke}, \citenamefont
  {Hosseini}, \citenamefont {Huyse}, \citenamefont {K{\"o}ster}, \citenamefont
  {Kudryavtsev}, \citenamefont {Laatiaoui}, \citenamefont {Lica}, \citenamefont
  {Magchiels}, \citenamefont {Manea}, \citenamefont {Merckling}, \citenamefont
  {Pereira}, \citenamefont {Raeder}, \citenamefont {Schumm}, \citenamefont
  {Sels}, \citenamefont {Thirolf}, \citenamefont {Tunhuma}, \citenamefont {Van
  Den~Bergh}, \citenamefont {Van~Duppen}, \citenamefont {Vantomme},
  \citenamefont {Verlinde}, \citenamefont {Villarreal},\ and\ \citenamefont
  {Wahl}}]{Kraemer2023}  \BibitemOpen
  \bibfield  {author} {\bibinfo {author} {\bibfnamefont {S.}~\bibnamefont
  {Kraemer}}, \bibinfo {author} {\bibfnamefont {J.}~\bibnamefont {Moens}},
  \bibinfo {author} {\bibfnamefont {M.}~\bibnamefont
  {Athanasakis-Kaklamanakis}}, \bibinfo {author} {\bibfnamefont
  {S.}~\bibnamefont {Bara}}, \bibinfo {author} {\bibfnamefont {K.}~\bibnamefont
  {Beeks}}, \bibinfo {author} {\bibfnamefont {P.}~\bibnamefont {Chhetri}},
  \bibinfo {author} {\bibfnamefont {K.}~\bibnamefont {Chrysalidis}}, \bibinfo
  {author} {\bibfnamefont {A.}~\bibnamefont {Claessens}}, \bibinfo {author}
  {\bibfnamefont {T.~E.}\ \bibnamefont {Cocolios}}, \bibinfo {author}
  {\bibfnamefont {J.~G.~M.}\ \bibnamefont {Correia}}, \bibinfo {author}
  {\bibfnamefont {H.}~\bibnamefont {De~Witte}}, \bibinfo {author}
  {\bibfnamefont {R.}~\bibnamefont {Ferrer}}, \bibinfo {author} {\bibfnamefont
  {S.}~\bibnamefont {Geldhof}}, \bibinfo {author} {\bibfnamefont
  {R.}~\bibnamefont {Heinke}}, \bibinfo {author} {\bibfnamefont
  {N.}~\bibnamefont {Hosseini}}, \bibinfo {author} {\bibfnamefont
  {M.}~\bibnamefont {Huyse}}, \bibinfo {author} {\bibfnamefont
  {U.}~\bibnamefont {K{\"o}ster}}, \bibinfo {author} {\bibfnamefont
  {Y.}~\bibnamefont {Kudryavtsev}}, \bibinfo {author} {\bibfnamefont
  {M.}~\bibnamefont {Laatiaoui}}, \bibinfo {author} {\bibfnamefont
  {R.}~\bibnamefont {Lica}}, \bibinfo {author} {\bibfnamefont {G.}~\bibnamefont
  {Magchiels}}, \bibinfo {author} {\bibfnamefont {V.}~\bibnamefont {Manea}},
  \bibinfo {author} {\bibfnamefont {C.}~\bibnamefont {Merckling}}, \bibinfo
  {author} {\bibfnamefont {L.~M.~C.}\ \bibnamefont {Pereira}}, \bibinfo
  {author} {\bibfnamefont {S.}~\bibnamefont {Raeder}}, \bibinfo {author}
  {\bibfnamefont {T.}~\bibnamefont {Schumm}}, \bibinfo {author} {\bibfnamefont
  {S.}~\bibnamefont {Sels}}, \bibinfo {author} {\bibfnamefont {P.~G.}\
  \bibnamefont {Thirolf}}, \bibinfo {author} {\bibfnamefont {S.~M.}\
  \bibnamefont {Tunhuma}}, \bibinfo {author} {\bibfnamefont {P.}~\bibnamefont
  {Van Den~Bergh}}, \bibinfo {author} {\bibfnamefont {P.}~\bibnamefont
  {Van~Duppen}}, \bibinfo {author} {\bibfnamefont {A.}~\bibnamefont
  {Vantomme}}, \bibinfo {author} {\bibfnamefont {M.}~\bibnamefont {Verlinde}},
  \bibinfo {author} {\bibfnamefont {R.}~\bibnamefont {Villarreal}},\ and\
  \bibinfo {author} {\bibfnamefont {U.}~\bibnamefont {Wahl}},\ }\bibfield
  {title} {\bibinfo {title} {Observation of the radiative decay of the
  {$^{229}$Th} nuclear clock isomer},\ }\href
  {https://doi.org/10.1038/s41586-023-05894-z} {\bibfield  {journal} {\bibinfo
  {journal} {Nature}\ }\textbf {\bibinfo {volume} {617}}(\bibinfo {number}
  {7962}),\ \bibinfo {pages} {706--710} (\bibinfo {year} {2023})}\BibitemShut
  {NoStop}\bibitem [{\citenamefont {Tkalya}\ \emph {et~al.}(2015)\citenamefont {Tkalya},
  \citenamefont {Schneider}, \citenamefont {Jeet},\ and\ \citenamefont
  {Hudson}}]{Tkalya2015}  \BibitemOpen
  \bibfield  {author} {\bibinfo {author} {\bibfnamefont {E.~V.}\ \bibnamefont
  {Tkalya}}, \bibinfo {author} {\bibfnamefont {C.}~\bibnamefont {Schneider}},
  \bibinfo {author} {\bibfnamefont {J.}~\bibnamefont {Jeet}},\ and\ \bibinfo
  {author} {\bibfnamefont {E.~R.}\ \bibnamefont {Hudson}},\ }\bibfield  {title}
  {\bibinfo {title} {Radiative lifetime and energy of the low-energy isomeric
  level in {$^{229}$Th}},\ }\href {https://doi.org/10.1103/physrevc.92.054324}
  {\bibfield  {journal} {\bibinfo  {journal} {Phys. Rev. C}\ }\textbf {\bibinfo
  {volume} {92}}(\bibinfo {number} {5}),\ \bibinfo {pages} {054324} (\bibinfo
  {year} {2015})}\BibitemShut {NoStop}\bibitem [{\citenamefont {Tkalya}\ \emph {et~al.}(2017)\citenamefont {Tkalya},
  \citenamefont {Schneider}, \citenamefont {Jeet},\ and\ \citenamefont
  {Hudson}}]{TkalyaErratum2017}  \BibitemOpen
  \bibfield  {author} {\bibinfo {author} {\bibfnamefont {E.~V.}\ \bibnamefont
  {Tkalya}}, \bibinfo {author} {\bibfnamefont {C.}~\bibnamefont {Schneider}},
  \bibinfo {author} {\bibfnamefont {J.}~\bibnamefont {Jeet}},\ and\ \bibinfo
  {author} {\bibfnamefont {E.~R.}\ \bibnamefont {Hudson}},\ }\bibfield  {title}
  {\bibinfo {title} {Erratum: Radiative lifetime and energy of the low-energy
  isomeric level in {$^{229}$Th} [{Phys. Rev. C} {92}, 054324 (2015)]},\ }\href
  {https://doi.org/10.1103/physrevc.95.039902} {\bibfield  {journal} {\bibinfo
  {journal} {Phys. Rev. C}\ }\textbf {\bibinfo {volume} {95}}(\bibinfo {number}
  {3}),\ \bibinfo {pages} {039902} (\bibinfo {year} {2017})}\BibitemShut
  {NoStop}\bibitem [{\citenamefont {Zhang}\ \emph {et~al.}(2022)\citenamefont {Zhang},
  \citenamefont {Wang},\ and\ \citenamefont {Wang}}]{Zhang2022}  \BibitemOpen
  \bibfield  {author} {\bibinfo {author} {\bibfnamefont {H.}~\bibnamefont
  {Zhang}}, \bibinfo {author} {\bibfnamefont {W.}~\bibnamefont {Wang}},\ and\
  \bibinfo {author} {\bibfnamefont {X.}~\bibnamefont {Wang}},\ }\bibfield
  {title} {\bibinfo {title} {Nuclear excitation cross section of {$^{229}$Th}
  via inelastic electron scattering},\ }\href
  {https://doi.org/10.1103/PhysRevC.106.044604} {\bibfield  {journal} {\bibinfo
   {journal} {Phys. Rev. C}\ }\textbf {\bibinfo {volume} {106}}(\bibinfo
  {number} {4}),\ \bibinfo {pages} {044604} (\bibinfo {year}
  {2022})}\BibitemShut {NoStop}\bibitem [{\citenamefont {Chen}\ \emph {et~al.}(2025)\citenamefont {Chen},
  \citenamefont {Wang},\ and\ \citenamefont {Wu}}]{Chen2025PSM}  \BibitemOpen
  \bibfield  {author} {\bibinfo {author} {\bibfnamefont {Z.-R.}\ \bibnamefont
  {Chen}}, \bibinfo {author} {\bibfnamefont {L.-J.}\ \bibnamefont {Wang}},\
  and\ \bibinfo {author} {\bibfnamefont {Y.}~\bibnamefont {Wu}},\ }\bibfield
  {title} {\bibinfo {title} {Microscopic nuclear structure study of
  {$^{229}$Th} by projected shell model},\ }\href
  {https://doi.org/10.1016/j.physletb.2025.139858} {\bibfield  {journal}
  {\bibinfo  {journal} {Phys. Lett. B}\ }\textbf {\bibinfo {volume} {869}},\
  \bibinfo {pages} {139858} (\bibinfo {year} {2025})}\BibitemShut {NoStop}\bibitem [{\citenamefont {Krane}(1974)}]{Krane1974}  \BibitemOpen
  \bibfield  {author} {\bibinfo {author} {\bibfnamefont {K.~S.}\ \bibnamefont
  {Krane}},\ }\bibfield  {title} {\bibinfo {title} {{$E2/M1$} multipole mixing
  ratios of {$2^{\prime}\rightarrow 2$} gamma transitions in even-even
  spherical nuclei},\ }\href {https://doi.org/10.1103/PhysRevC.10.1197}
  {\bibfield  {journal} {\bibinfo  {journal} {Phys. Rev. C}\ }\textbf {\bibinfo
  {volume} {10}}(\bibinfo {number} {3}),\ \bibinfo {pages} {1197--1210}
  (\bibinfo {year} {1974})}\BibitemShut {NoStop}\bibitem [{\citenamefont {Smith}\ \emph {et~al.}(2019)\citenamefont {Smith},
  \citenamefont {MacLean}, \citenamefont {Ashfield}, \citenamefont {Chester},
  \citenamefont {Garnsworthy},\ and\ \citenamefont
  {Svensson}}]{Smith2019GRIFFIN}  \BibitemOpen
  \bibfield  {author} {\bibinfo {author} {\bibfnamefont {J.}~\bibnamefont
  {Smith}}, \bibinfo {author} {\bibfnamefont {A.}~\bibnamefont {MacLean}},
  \bibinfo {author} {\bibfnamefont {W.}~\bibnamefont {Ashfield}}, \bibinfo
  {author} {\bibfnamefont {A.}~\bibnamefont {Chester}}, \bibinfo {author}
  {\bibfnamefont {A.}~\bibnamefont {Garnsworthy}},\ and\ \bibinfo {author}
  {\bibfnamefont {C.}~\bibnamefont {Svensson}},\ }\bibfield  {title} {\bibinfo
  {title} {Gamma--gamma angular correlation analysis techniques with the
  {GRIFFIN} spectrometer},\ }\href {https://doi.org/10.1016/j.nima.2018.10.097}
  {\bibfield  {journal} {\bibinfo  {journal} {Nucl. Instrum. Methods Phys. Res.
  A}\ }\textbf {\bibinfo {volume} {922}},\ \bibinfo {pages} {47--63} (\bibinfo
  {year} {2019})}\BibitemShut {NoStop}\bibitem [{\citenamefont {Porsev}\ and\ \citenamefont
  {Flambaum}(2010)}]{Porsev2010Th3}  \BibitemOpen
  \bibfield  {author} {\bibinfo {author} {\bibfnamefont {S.~G.}\ \bibnamefont
  {Porsev}}\ and\ \bibinfo {author} {\bibfnamefont {V.~V.}\ \bibnamefont
  {Flambaum}},\ }\bibfield  {title} {\bibinfo {title} {Effect of atomic
  electrons on the 7.6-ev nuclear transition in {$^{229}$Th$^{3+}$}},\ }\href
  {https://doi.org/10.1103/PhysRevA.81.032504} {\bibfield  {journal} {\bibinfo
  {journal} {Phys. Rev. A}\ }\textbf {\bibinfo {volume} {81}}(\bibinfo {number}
  {3}),\ \bibinfo {pages} {032504} (\bibinfo {year} {2010})}\BibitemShut
  {NoStop}\bibitem [{\citenamefont {Porsev}\ \emph {et~al.}(2010)\citenamefont {Porsev},
  \citenamefont {Flambaum}, \citenamefont {Peik},\ and\ \citenamefont
  {Tamm}}]{PorsevBridge2010}  \BibitemOpen
  \bibfield  {author} {\bibinfo {author} {\bibfnamefont {S.~G.}\ \bibnamefont
  {Porsev}}, \bibinfo {author} {\bibfnamefont {V.~V.}\ \bibnamefont
  {Flambaum}}, \bibinfo {author} {\bibfnamefont {E.}~\bibnamefont {Peik}},\
  and\ \bibinfo {author} {\bibfnamefont {C.}~\bibnamefont {Tamm}},\ }\bibfield
  {title} {\bibinfo {title} {Excitation of the isomeric {$^{229m}$Th} nuclear
  state via an electronic bridge process in {$^{229}$Th$^+$}},\ }\href
  {https://doi.org/10.1103/physrevlett.105.182501} {\bibfield  {journal}
  {\bibinfo  {journal} {Phys. Rev. Lett.}\ }\textbf {\bibinfo {volume}
  {105}}(\bibinfo {number} {18}),\ \bibinfo {pages} {182501} (\bibinfo {year}
  {2010})}\BibitemShut {NoStop}\bibitem [{\citenamefont {Wang}\ \emph {et~al.}(2024)\citenamefont {Wang},
  \citenamefont {Zou}, \citenamefont {Fritzsche},\ and\ \citenamefont
  {Li}}]{Wang2024}  \BibitemOpen
  \bibfield  {author} {\bibinfo {author} {\bibfnamefont {W.}~\bibnamefont
  {Wang}}, \bibinfo {author} {\bibfnamefont {F.}~\bibnamefont {Zou}}, \bibinfo
  {author} {\bibfnamefont {S.}~\bibnamefont {Fritzsche}},\ and\ \bibinfo
  {author} {\bibfnamefont {Y.}~\bibnamefont {Li}},\ }\bibfield  {title}
  {\bibinfo {title} {Isomeric population transfer of the {$^{229}$Th} nucleus
  via hyperfine electronic bridge},\ }\href
  {https://doi.org/10.1103/PhysRevLett.133.223001} {\bibfield  {journal}
  {\bibinfo  {journal} {Phys. Rev. Lett.}\ }\textbf {\bibinfo {volume}
  {133}}(\bibinfo {number} {22}),\ \bibinfo {pages} {223001} (\bibinfo {year}
  {2024})}\BibitemShut {NoStop}\bibitem [{\citenamefont {Blaise}\ and\ \citenamefont
  {Wyart}(1992)}]{Blaise2021}  \BibitemOpen
  \bibfield  {author} {\bibinfo {author} {\bibfnamefont {J.}~\bibnamefont
  {Blaise}}\ and\ \bibinfo {author} {\bibfnamefont {J.-F.}\ \bibnamefont
  {Wyart}},\ }\href {http://www.lac.universite-paris-saclay.fr/Data/Database/}
  {\bibinfo {title} {Energy levels and atomic spectra of actinides}},\ \bibinfo
  {howpublished} {International Tables of Selected Constants, Vol. 20 (Paris)}
  (\bibinfo {year} {1992})\BibitemShut {NoStop}\bibitem [{\citenamefont {Li}\ \emph {et~al.}(2022)\citenamefont {Li},
  \citenamefont {Qiao}, \citenamefont {Tang},\ and\ \citenamefont
  {Shi}}]{Li2022}  \BibitemOpen
  \bibfield  {author} {\bibinfo {author} {\bibfnamefont {F.-C.}\ \bibnamefont
  {Li}}, \bibinfo {author} {\bibfnamefont {H.-X.}\ \bibnamefont {Qiao}},
  \bibinfo {author} {\bibfnamefont {Y.-B.}\ \bibnamefont {Tang}},\ and\
  \bibinfo {author} {\bibfnamefont {T.-Y.}\ \bibnamefont {Shi}},\ }\bibfield
  {title} {\bibinfo {title} {Relativistic coupled-cluster calculations of the
  energies, {Land{\'e}} g factors, radiative transition properties including
  line strengths, oscillator strengths, and transition rates for {Th}$^{3+}$},\
  }\href {https://doi.org/10.1016/j.jqsrt.2022.108241} {\bibfield  {journal}
  {\bibinfo  {journal} {J. Quant. Spectrosc. Radiat. Transfer}\ }\textbf
  {\bibinfo {volume} {288}},\ \bibinfo {pages} {108241} (\bibinfo {year}
  {2022})}\BibitemShut {NoStop}\bibitem [{\citenamefont {Cheung}\ \emph {et~al.}(2025)\citenamefont {Cheung},
  \citenamefont {Kozlov}, \citenamefont {Porsev}, \citenamefont {Safronova},
  \citenamefont {Tupitsyn},\ and\ \citenamefont {Bondarev}}]{Cheung2025PCI}  \BibitemOpen
  \bibfield  {author} {\bibinfo {author} {\bibfnamefont {C.}~\bibnamefont
  {Cheung}}, \bibinfo {author} {\bibfnamefont {M.~G.}\ \bibnamefont {Kozlov}},
  \bibinfo {author} {\bibfnamefont {S.~G.}\ \bibnamefont {Porsev}}, \bibinfo
  {author} {\bibfnamefont {M.~S.}\ \bibnamefont {Safronova}}, \bibinfo {author}
  {\bibfnamefont {I.~I.}\ \bibnamefont {Tupitsyn}},\ and\ \bibinfo {author}
  {\bibfnamefont {A.~I.}\ \bibnamefont {Bondarev}},\ }\bibfield  {title}
  {\bibinfo {title} {{pCI}: A parallel configuration interaction software
  package for high-precision atomic structure calculations},\ }\href
  {https://doi.org/10.1016/j.cpc.2024.109463} {\bibfield  {journal} {\bibinfo
  {journal} {Comput. Phys. Commun.}\ }\textbf {\bibinfo {volume} {308}},\
  \bibinfo {pages} {109463} (\bibinfo {year} {2025})}\BibitemShut {NoStop}\bibitem [{\citenamefont {Kahl}\ and\ \citenamefont
  {Berengut}(2019)}]{Kahl2019}  \BibitemOpen
  \bibfield  {author} {\bibinfo {author} {\bibfnamefont {E.~V.}\ \bibnamefont
  {Kahl}}\ and\ \bibinfo {author} {\bibfnamefont {J.~C.}\ \bibnamefont
  {Berengut}},\ }\bibfield  {title} {\bibinfo {title} {{AMBiT}: A programme for
  high-precision relativistic atomic structure calculations},\ }\href
  {https://doi.org/10.1016/j.cpc.2018.12.014} {\bibfield  {journal} {\bibinfo
  {journal} {Comput. Phys. Commun.}\ }\textbf {\bibinfo {volume} {238}},\
  \bibinfo {pages} {232--243} (\bibinfo {year} {2019})}\BibitemShut {NoStop}\bibitem [{Sup()}]{Supplemental}  \BibitemOpen
  \href@noop {} {}\bibinfo {note} {See the Supplemental Material appended below for calculation
  details (Sec. I) and uncertainty analysis (Sec. II)}\BibitemShut {NoStop}\bibitem [{\citenamefont {Radnaev}\ \emph {et~al.}(2012)\citenamefont
  {Radnaev}, \citenamefont {Campbell},\ and\ \citenamefont
  {Kuzmich}}]{Radnaev2012}  \BibitemOpen
  \bibfield  {author} {\bibinfo {author} {\bibfnamefont {A.~G.}\ \bibnamefont
  {Radnaev}}, \bibinfo {author} {\bibfnamefont {C.~J.}\ \bibnamefont
  {Campbell}},\ and\ \bibinfo {author} {\bibfnamefont {A.}~\bibnamefont
  {Kuzmich}},\ }\bibfield  {title} {\bibinfo {title} {Observation of the 717-nm
  electric quadrupole transition in triply charged thorium},\ }\href
  {https://doi.org/10.1103/PhysRevA.86.060501} {\bibfield  {journal} {\bibinfo
  {journal} {Phys. Rev. A}\ }\textbf {\bibinfo {volume} {86}}(\bibinfo {number}
  {6}),\ \bibinfo {pages} {060501(R)} (\bibinfo {year} {2012})}\BibitemShut
  {NoStop}\bibitem [{\citenamefont {Wang}\ \emph {et~al.}(2022)\citenamefont {Wang},
  \citenamefont {Yang}, \citenamefont {Wang}, \citenamefont {Zhang},
  \citenamefont {Jia}, \citenamefont {Zhu}, \citenamefont {Yan},\ and\
  \citenamefont {Yang}}]{Wang2022Ratio}  \BibitemOpen
  \bibfield  {author} {\bibinfo {author} {\bibfnamefont {G.}~\bibnamefont
  {Wang}}, \bibinfo {author} {\bibfnamefont {M.}~\bibnamefont {Yang}}, \bibinfo
  {author} {\bibfnamefont {E.}~\bibnamefont {Wang}}, \bibinfo {author}
  {\bibfnamefont {X.}~\bibnamefont {Zhang}}, \bibinfo {author} {\bibfnamefont
  {A.}~\bibnamefont {Jia}}, \bibinfo {author} {\bibfnamefont {L.}~\bibnamefont
  {Zhu}}, \bibinfo {author} {\bibfnamefont {S.}~\bibnamefont {Yan}},\ and\
  \bibinfo {author} {\bibfnamefont {J.}~\bibnamefont {Yang}},\ }\bibfield
  {title} {\bibinfo {title} {In situ probing and stabilizing the power ratio of
  electro-optic-modulated laser pairs based on {VIPA} etalon for quantum
  sensing},\ }\href {https://doi.org/10.1364/OL.458844} {\bibfield  {journal}
  {\bibinfo  {journal} {Opt. Lett.}\ }\textbf {\bibinfo {volume} {47}}(\bibinfo
  {number} {12}),\ \bibinfo {pages} {2983--2986} (\bibinfo {year}
  {2022})}\BibitemShut {NoStop}\bibitem [{\citenamefont {Kaneda}\ \emph {et~al.}(2016)\citenamefont {Kaneda},
  \citenamefont {Yarborough}, \citenamefont {Merzlyak}, \citenamefont
  {Yamaguchi}, \citenamefont {Hayashida}, \citenamefont {Ohmae},\ and\
  \citenamefont {Katori}}]{Kaneda2016}  \BibitemOpen
  \bibfield  {author} {\bibinfo {author} {\bibfnamefont {Y.}~\bibnamefont
  {Kaneda}}, \bibinfo {author} {\bibfnamefont {J.~M.}\ \bibnamefont
  {Yarborough}}, \bibinfo {author} {\bibfnamefont {Y.}~\bibnamefont
  {Merzlyak}}, \bibinfo {author} {\bibfnamefont {A.}~\bibnamefont {Yamaguchi}},
  \bibinfo {author} {\bibfnamefont {K.}~\bibnamefont {Hayashida}}, \bibinfo
  {author} {\bibfnamefont {N.}~\bibnamefont {Ohmae}},\ and\ \bibinfo {author}
  {\bibfnamefont {H.}~\bibnamefont {Katori}},\ }\bibfield  {title} {\bibinfo
  {title} {Continuous-wave, single-frequency 229 nm laser source for laser
  cooling of cadmium atoms},\ }\href {https://doi.org/10.1364/OL.41.000705}
  {\bibfield  {journal} {\bibinfo  {journal} {Opt. Lett.}\ }\textbf {\bibinfo
  {volume} {41}}(\bibinfo {number} {4}),\ \bibinfo {pages} {705--708} (\bibinfo
  {year} {2016})}\BibitemShut {NoStop}\bibitem [{\citenamefont {Preuschoff}\ \emph {et~al.}(2022)\citenamefont
  {Preuschoff}, \citenamefont {Baus}, \citenamefont {Schlosser},\ and\
  \citenamefont {Birkl}}]{Preuschoff2022}  \BibitemOpen
  \bibfield  {author} {\bibinfo {author} {\bibfnamefont {T.}~\bibnamefont
  {Preuschoff}}, \bibinfo {author} {\bibfnamefont {P.}~\bibnamefont {Baus}},
  \bibinfo {author} {\bibfnamefont {M.}~\bibnamefont {Schlosser}},\ and\
  \bibinfo {author} {\bibfnamefont {G.}~\bibnamefont {Birkl}},\ }\bibfield
  {title} {\bibinfo {title} {Wideband current modulation of diode lasers for
  frequency stabilization},\ }\href {https://doi.org/10.1063/5.0093520}
  {\bibfield  {journal} {\bibinfo  {journal} {Rev. Sci. Instrum.}\ }\textbf
  {\bibinfo {volume} {93}},\ \bibinfo {pages} {063002} (\bibinfo {year}
  {2022})}\BibitemShut {NoStop}\bibitem [{\citenamefont {Bilous}\ \emph {et~al.}(2020)\citenamefont {Bilous},
  \citenamefont {Bekker}, \citenamefont {Berengut}, \citenamefont {Seiferle},
  \citenamefont {von~der Wense}, \citenamefont {Thirolf}, \citenamefont
  {Pfeifer}, \citenamefont {Crespo L{\'o}pez-Urrutia},\ and\ \citenamefont
  {P{\'a}lffy}}]{Bilous2020}  \BibitemOpen
  \bibfield  {author} {\bibinfo {author} {\bibfnamefont {P.~V.}\ \bibnamefont
  {Bilous}}, \bibinfo {author} {\bibfnamefont {H.}~\bibnamefont {Bekker}},
  \bibinfo {author} {\bibfnamefont {J.~C.}\ \bibnamefont {Berengut}}, \bibinfo
  {author} {\bibfnamefont {B.}~\bibnamefont {Seiferle}}, \bibinfo {author}
  {\bibfnamefont {L.}~\bibnamefont {von~der Wense}}, \bibinfo {author}
  {\bibfnamefont {P.~G.}\ \bibnamefont {Thirolf}}, \bibinfo {author}
  {\bibfnamefont {T.}~\bibnamefont {Pfeifer}}, \bibinfo {author} {\bibfnamefont
  {J.~R.}\ \bibnamefont {Crespo L{\'o}pez-Urrutia}},\ and\ \bibinfo {author}
  {\bibfnamefont {A.}~\bibnamefont {P{\'a}lffy}},\ }\bibfield  {title}
  {\bibinfo {title} {Electronic bridge excitation in highly charged
  {$^{229}$Th} ions},\ }\href {https://doi.org/10.1103/PhysRevLett.124.192502}
  {\bibfield  {journal} {\bibinfo  {journal} {Phys. Rev. Lett.}\ }\textbf
  {\bibinfo {volume} {124}}(\bibinfo {number} {19}),\ \bibinfo {pages} {192502}
  (\bibinfo {year} {2020})}\BibitemShut {NoStop}\bibitem [{\citenamefont {Maurya}\ \emph {et~al.}(2026)\citenamefont {Maurya},
  \citenamefont {Lal}, \citenamefont {Tiedau}, \citenamefont {Okhapkin},\ and\
  \citenamefont {Peik}}]{Maurya2026}  \BibitemOpen
  \bibfield  {author} {\bibinfo {author} {\bibfnamefont {S.~S.}\ \bibnamefont
  {Maurya}}, \bibinfo {author} {\bibfnamefont {V.}~\bibnamefont {Lal}},
  \bibinfo {author} {\bibfnamefont {J.}~\bibnamefont {Tiedau}}, \bibinfo
  {author} {\bibfnamefont {M.~V.}\ \bibnamefont {Okhapkin}},\ and\ \bibinfo
  {author} {\bibfnamefont {E.}~\bibnamefont {Peik}},\ }\bibfield  {title}
  {\bibinfo {title} {{$J=0$} metastable state of {Th$^{2+}$} for a
  hyperfine-free nuclear clock},\ }\href {https://doi.org/10.1103/c98g-6f84}
  {\bibfield  {journal} {\bibinfo  {journal} {Phys. Rev. A}\ }\textbf {\bibinfo
  {volume} {114}}(\bibinfo {number} {3}),\ \bibinfo {pages} {032819} (\bibinfo
  {year} {2026})}\BibitemShut {NoStop}\bibitem [{\citenamefont {Yu}\ \emph {et~al.}(2025)\citenamefont {Yu},
  \citenamefont {Hua}, \citenamefont {Tong}, \citenamefont {Li},\ and\
  \citenamefont {She}}]{Yu2025}  \BibitemOpen
  \bibfield  {author} {\bibinfo {author} {\bibfnamefont {S.-C.}\ \bibnamefont
  {Yu}}, \bibinfo {author} {\bibfnamefont {X.}~\bibnamefont {Hua}}, \bibinfo
  {author} {\bibfnamefont {X.}~\bibnamefont {Tong}}, \bibinfo {author}
  {\bibfnamefont {C.-B.}\ \bibnamefont {Li}},\ and\ \bibinfo {author}
  {\bibfnamefont {L.}~\bibnamefont {She}},\ }\bibfield  {title} {\bibinfo
  {title} {Highly charged {$^{229}$Th$^{6+}$} ions as the candidate platform
  for nuclear clock},\ }\href {https://doi.org/10.1088/0256-307X/42/12/120302}
  {\bibfield  {journal} {\bibinfo  {journal} {Chin. Phys. Lett.}\ }\textbf
  {\bibinfo {volume} {42}}(\bibinfo {number} {12}),\ \bibinfo {pages} {120302}
  (\bibinfo {year} {2025})}\BibitemShut {NoStop}\bibitem [{\citenamefont {Campbell}\ \emph {et~al.}(2011)\citenamefont
  {Campbell}, \citenamefont {Radnaev},\ and\ \citenamefont
  {Kuzmich}}]{Campbell2011}  \BibitemOpen
  \bibfield  {author} {\bibinfo {author} {\bibfnamefont {C.~J.}\ \bibnamefont
  {Campbell}}, \bibinfo {author} {\bibfnamefont {A.~G.}\ \bibnamefont
  {Radnaev}},\ and\ \bibinfo {author} {\bibfnamefont {A.}~\bibnamefont
  {Kuzmich}},\ }\bibfield  {title} {\bibinfo {title} {Wigner crystals of
  {$^{229}$Th} for optical excitation of the nuclear isomer},\ }\href
  {https://doi.org/10.1103/PhysRevLett.106.223001} {\bibfield  {journal}
  {\bibinfo  {journal} {Phys. Rev. Lett.}\ }\textbf {\bibinfo {volume}
  {106}}(\bibinfo {number} {22}),\ \bibinfo {pages} {223001} (\bibinfo {year}
  {2011})}\BibitemShut {NoStop}\bibitem [{\citenamefont {Porsev}\ \emph {et~al.}(2021)\citenamefont {Porsev},
  \citenamefont {Safronova},\ and\ \citenamefont {Kozlov}}]{Porsev2021}  \BibitemOpen
  \bibfield  {author} {\bibinfo {author} {\bibfnamefont {S.~G.}\ \bibnamefont
  {Porsev}}, \bibinfo {author} {\bibfnamefont {M.~S.}\ \bibnamefont
  {Safronova}},\ and\ \bibinfo {author} {\bibfnamefont {M.~G.}\ \bibnamefont
  {Kozlov}},\ }\bibfield  {title} {\bibinfo {title} {Precision calculation of
  hyperfine constants for extracting nuclear moments of {$^{229}$Th}},\ }\href
  {https://doi.org/10.1103/PhysRevLett.127.253001} {\bibfield  {journal}
  {\bibinfo  {journal} {Phys. Rev. Lett.}\ }\textbf {\bibinfo {volume}
  {127}}(\bibinfo {number} {25}),\ \bibinfo {pages} {253001} (\bibinfo {year}
  {2021})}\BibitemShut {NoStop}\bibitem [{\citenamefont {Yu}\ \emph {et~al.}(2024)\citenamefont {Yu},
  \citenamefont {Gan}, \citenamefont {Hua}, \citenamefont {Tong},\ and\
  \citenamefont {Li}}]{YuClock2024}  \BibitemOpen
  \bibfield  {author} {\bibinfo {author} {\bibfnamefont {S.-C.}\ \bibnamefont
  {Yu}}, \bibinfo {author} {\bibfnamefont {W.-T.}\ \bibnamefont {Gan}},
  \bibinfo {author} {\bibfnamefont {X.}~\bibnamefont {Hua}}, \bibinfo {author}
  {\bibfnamefont {X.}~\bibnamefont {Tong}},\ and\ \bibinfo {author}
  {\bibfnamefont {C.-B.}\ \bibnamefont {Li}},\ }\bibfield  {title} {\bibinfo
  {title} {{$^{229}$Th$^{3+}$} as an ionic optical clock for
  fine-structure-constant variations},\ }\href
  {https://doi.org/10.1103/physreva.109.063115} {\bibfield  {journal} {\bibinfo
   {journal} {Phys. Rev. A}\ }\textbf {\bibinfo {volume} {109}}(\bibinfo
  {number} {6}),\ \bibinfo {pages} {063115} (\bibinfo {year}
  {2024})}\BibitemShut {NoStop}\bibitem [{\citenamefont {Yu}\ \emph {et~al.}(2026)\citenamefont {Yu},
  \citenamefont {Li},\ and\ \citenamefont {She}}]{YuBridge2026}  \BibitemOpen
  \bibfield  {author} {\bibinfo {author} {\bibfnamefont {S.-C.}\ \bibnamefont
  {Yu}}, \bibinfo {author} {\bibfnamefont {C.-B.}\ \bibnamefont {Li}},\ and\
  \bibinfo {author} {\bibfnamefont {L.}~\bibnamefont {She}},\ }\bibfield
  {title} {\bibinfo {title} {Prediction of {$I_g,6d_{3/2} \to I_g,7p_{1/2}$},
  {$I_g,7s_{1/2} \to I_g,7p_{1/2}$} and {$I_g,7p_{1/2} \to I_m,7s_{1/2}$}
  transition frequencies in {$^{229}$Th$^{3+}$} ion},\ }\href
  {https://doi.org/10.1088/1674-1056/ae0016} {\bibfield  {journal} {\bibinfo
  {journal} {Chin. Phys. B}\ }\textbf {\bibinfo {volume} {35}}(\bibinfo
  {number} {2}),\ \bibinfo {pages} {023701} (\bibinfo {year}
  {2026})}\BibitemShut {NoStop}\bibitem [{\citenamefont {Porsev}\ and\ \citenamefont
  {Safronova}(2026)}]{Porsev2026}  \BibitemOpen
  \bibfield  {author} {\bibinfo {author} {\bibfnamefont {S.~G.}\ \bibnamefont
  {Porsev}}\ and\ \bibinfo {author} {\bibfnamefont {M.~S.}\ \bibnamefont
  {Safronova}},\ }\href {https://doi.org/10.48550/arXiv.2608.13700} {\bibinfo
  {title} {Precision calculations of the hyperfine constants and isotope and
  isomer shifts in low-lying states of {$^{229}$Th$^{3+}$}}} (\bibinfo {year}
  {2026}),\ \Eprint {https://arxiv.org/abs/2608.13700} {arXiv:2608.13700
  [physics.atom-ph]} \BibitemShut {NoStop}\bibitem [{\citenamefont {Safronova}\ \emph {et~al.}(2013)\citenamefont
  {Safronova}, \citenamefont {Safronova}, \citenamefont {Radnaev},
  \citenamefont {Campbell},\ and\ \citenamefont {Kuzmich}}]{Safronova2013}  \BibitemOpen
  \bibfield  {author} {\bibinfo {author} {\bibfnamefont {M.~S.}\ \bibnamefont
  {Safronova}}, \bibinfo {author} {\bibfnamefont {U.~I.}\ \bibnamefont
  {Safronova}}, \bibinfo {author} {\bibfnamefont {A.~G.}\ \bibnamefont
  {Radnaev}}, \bibinfo {author} {\bibfnamefont {C.~J.}\ \bibnamefont
  {Campbell}},\ and\ \bibinfo {author} {\bibfnamefont {A.}~\bibnamefont
  {Kuzmich}},\ }\bibfield  {title} {\bibinfo {title} {Magnetic dipole and
  electric quadrupole moments of the {$^{229}$Th} nucleus},\ }\href
  {https://doi.org/10.1103/PhysRevA.88.060501} {\bibfield  {journal} {\bibinfo
  {journal} {Phys. Rev. A}\ }\textbf {\bibinfo {volume} {88}}(\bibinfo {number}
  {6}),\ \bibinfo {pages} {060501(R)} (\bibinfo {year} {2013})}\BibitemShut
  {NoStop}\end{thebibliography}
\end{document}